\pdfoutput=1
\documentclass{ieeeaccess}
\usepackage{cite}
\usepackage{amsmath,amssymb,amsfonts}
\usepackage{algorithmic}
\usepackage{graphicx}
\usepackage{textcomp}
\usepackage{cite}
\usepackage{amsmath,amssymb,amsfonts}
\usepackage{algorithmic}
\usepackage{graphicx,color}
\usepackage{textcomp}
\usepackage{xcolor}
\usepackage{hyperref}
\usepackage{multirow}
\usepackage{lipsum}
\usepackage{algorithm}
\usepackage{algorithmic}
\usepackage{subcaption}
\usepackage{paralist}
\usepackage{booktabs} 
\usepackage{adjustbox} 
\usepackage{longtable} 
\usepackage{multicol} 
\usepackage{graphicx}
\usepackage{multirow}
\usepackage{booktabs}
\usepackage{xcolor}
\usepackage{colortbl}
\usepackage{subcaption}
\usepackage{paralist}

\usepackage{xcolor}
\usepackage{colortbl}
\usepackage{pifont}  

\definecolor{headerblue}{RGB}{31,78,121}
\definecolor{lightgreen}{RGB}{198,239,206}
\definecolor{lightred}{RGB}{255,199,206}
\definecolor{lightyellow}{RGB}{255,235,156}
\definecolor{recorange}{RGB}{237,125,49}

\def\BibTeX{{\rm B\kern-.05em{\sc i\kern-.025em b}\kern-.08em
    T\kern-.1667em\lower.7ex\hbox{E}\kern-.125emX}}
\begin{document}
\doi{}

\title{Categorical Robustness Assessment for Machine Learning based Network Intrusion Detection Systems}
\author{Mayank Raj\authorrefmark{1}, Nathaniel D. Bastian \authorrefmark{2}, Lance Fiondella\authorrefmark{1},  and Gokhan Kul\authorrefmark{1}
\\\authorrefmark{1}University of Massachusetts Dartmouth, Dartmouth, MA 02747 USA\\
\authorrefmark{2}United States Military Academy, West Point, NY 10996 USA
\corresp{Corresponding author: Mayank Raj (email: mraj1@umassd.edu).}}

\markboth
{Raj \headeretal: Categorical Robustness Assessment for ML-based Network Intrusion Detection Systems}
{Raj \headeretal: Categorical Robustness Assessment for ML-based Network Intrusion Detection Systems}


\begin{abstract}
Network Intrusion Detection Systems (NIDS) heavily utlize Machine Learning (ML) but ML models can be manipulated via adversarial attacks. These attacks add carefully crafted perturbations to network traffic data that leads to misclassifications. While prior work has demonstrated adversarial vulnerabilities in isolated settings, systematic cross-architecture  as well as class and category of attack based comparisons under controlled attack conditions remain limited, leaving practitioners without clear guidance on which models to deploy in adversarial environments. This paper asks a simple question: what type of classifier architectures actually hold up when attackers try to manipulate the systems? We put three popular architectures through their paces: a 1D Convolutional Neural Network, a Long Short-Term Memory (LSTM) network, and a Random Forest (RF) ensemble. Using the ACI-IoT-2023 dataset (over 1.2 million samples spanning 12 attack types), we subject each model with FGSM and PGD adversarial attacks, which apply gradient-based perturbations in normalized feature space consistent with established adversarial ML evaluation protocols, at perturbation budgets ranging from $\epsilon=0.01$ to $\epsilon=0.1$. Surprisingly, Random Forest achieved near-perfect baseline accuracy (99.98\%), yet collapsed catastrophically under attack, dropping 73 percentage points at the smallest perturbation we tested. CNN, on the other hand, retained 95.5\% accuracy at $\epsilon=0.01$ and degraded gracefully as perturbations increased. LSTM fell somewhere in between. These findings flip the conventional wisdom where high baseline accuracy means nothing if a model shatters at the first sign of adversarial pressure. For practitioners deploying intrusion detection in adversarial environments, we recommend CNN-based architectures and provide scenario-specific deployment guidance. We also identify class imbalance as a compounding vulnerability due to the fact that minority attack classes struggle at baseline became completely undetectable under attack in many scenarios.

\end{abstract}

\begin{IEEEkeywords}
%
Adversarial machine learning, network intrusion detection, deep learning, random forest, CNN, LSTM, FGSM, PGD, IoT security, robustness evaluation
\end{IEEEkeywords}


\maketitle

\makeatletter
\renewcommand{\ps@headings}{%
  \let\@oddhead\@empty
  \let\@evenhead\@empty
  \let\@oddfoot\@empty
  \let\@evenfoot\@empty
}
\pagestyle{headings}
\thispagestyle{empty}
\makeatother

\section{INTRODUCTION}
\label{sec:1intro}
\IEEEPARstart{M}{achine} learning–based Network Intrusion Detection Systems (NIDS) have achieved near-perfect classification accuracy on benchmark datasets \cite{ref1,ref2,ref3}. Yet this reported success conceals a critical vulnerability: these systems catastrophically fail under adversarial perturbations \cite{ref4,ref5,ref6}. This paper introduces the \textit{False Champion Problem}, wherein models with the highest baseline accuracy suffer the most severe degradation under adversarial attack, fundamentally challenging conventional NIDS evaluation paradigms.

Internet of Things (IoT) devices has significantly expanded the global attack surface \cite{ref7,ref8}. With over 15 billion connected devices in 2025, and the increasing sophistication of Advanced Persistent Threats (APTs), maintaining reliable automated intrusion detection under adversarial conditions is now mission-critical \cite{ref9}. In this threat landscape, an NIDS achieving 99.98\% baseline accuracy but collapsing to 26\% under minimal perturbation provides a dangerously misleading sense of security, which may be more harmful than having no detection capability at all.

Prior research has demonstrated that adversarial attacks against ML-based NIDS are both feasible and highly effective \cite{ref10,ref11,ref12,ref13}. Carefully crafted perturbations allow malicious traffic to evade detection while preserving functional semantics \cite{ref14,ref15}. However, one fundamental question remains unanswered: \textit{what architectural properties make certain models brittle or resilient under adversarial manipulation?} Understanding this relationship is essential for designing NIDS that maintain operational effectiveness in contested environments where adversaries actively attempt to evade detection.

This work provides an architectural analysis of adversarial robustness across three fundamentally distinct machine learning paradigms: Convolutional Neural Networks (CNNs) \cite{ref16,ref17}, Long Short-Term Memory networks (LSTMs) \cite{ref18,ref19}, and Random Forests \cite{ref20,ref21}. Our findings show that the very mechanisms enabling high baseline accuracy in tree-based ensembles, namely axis-aligned recursive feature partitioning are the precise properties that render these models catastrophically vulnerable to adversarial perturbations. The threshold-based splits create discontinuous decision boundaries that adversaries easily exploit. In contrast, neural architectures that learn distributed and continuous representations through gradient-based optimization exhibit inherent regularization properties conferring significantly greater robustness, even when their baseline accuracy is marginally lower. The smooth, continuous decision surfaces of CNNs and LSTMs require substantially larger perturbations to cross, naturally increasing resistance to evasion.

We evaluate model robustness using the ACI-IoT-2023 dataset \cite{ref22}, a comprehensive IoT network traffic collection comprising over 1.23 million samples across 12 attack classes organized into five categories: Benign, Reconnaissance (Recon), Denial-of-Service (DoS), Brute Force, and Spoofing. The dataset exhibits substantial real-world class imbalance. For example, Port Scan accounts for 35.8\% of all samples, while ARP Spoofing represents less than 0.01\%. This imbalance creates realistic challenges for robust classifier design and enables examination of how adversarial vulnerability correlates with class prevalence. The dataset classes and their distribution is shown on Table~\ref{tab:dataset}. 

\begin{table}[!t]
\caption{ACI-IoT Network Dataset Composition Before Preprocessing}
\label{tab:dataset}
\centering
\begin{tabular}{|l|l|r|r|}
\hline
\textbf{Category} & \textbf{Class} & \textbf{Count} & \textbf{\%} \\
\hline
Benign & Benign & 329,295 & 26.7 \\
\hline
\multirow{4}{*}{Recon} & Ping Sweep & 71,928 & 5.8 \\
& OS Scan & 37,524 & 3.0 \\
& Vulnerability Scan & 39,537 & 3.2 \\
& Port Scan & 441,282 & 35.8 \\
\hline
\multirow{5}{*}{DoS} & ICMP Flooding & 225,234 & 18.2 \\
& Slowloris & 18,643 & 1.5 \\
& SYN Flood & 13,857 & 1.1 \\
& UDP Flood & 791 & 0.06 \\
& DNS Flood & 46,935 & 3.8 \\
\hline
Brute Force & Dictionary Brute Force & 6,380 & 0.5 \\
\hline
Spoofing & ARP Spoofing & 5 & 0.0 \\
\hline
\multicolumn{2}{|l|}{\textbf{TOTAL}} & \textbf{1,231,411} & \textbf{100} \\
\hline
\end{tabular}
\end{table}

Our evaluation methodology, illustrated in Fig.~\ref{fig:framework}, integrates a multi-dimensional robustness assessment framework that extends beyond simple accuracy degradation. Specifically, we incorporate: (1) gradient-based adversarial attacks, including FGSM \cite{ref4} with perturbation magnitudes $\epsilon \in \{0.01, 0.05, 0.1\}$ and PGD \cite{ref5} with 40 iterations and step size $\alpha = \epsilon/4$; (2) transfer attacks using CNN-generated adversarial examples to attack non-differentiable Random Forests \cite{ref22,ref23}; (3) certified robustness metrics, including CLEVER L$_2$ and L$_\infty$ scores \cite{ref20}; (4) perturbability analysis to measure inter-class feature space distances; (5) distribution shift analysis including covariate (Gaussian noise) and label shift; and (6) diffusion-based reconstruction to assess model generalization capacity. This framework enables granular vulnerability analysis at both class and category levels, uncovering weaknesses that remain hidden when relying solely on aggregate metrics.

Our results reveal three critical findings with broad implications for NIDS deployment in adversarial environments. First, we empirically demonstrate the \textit{False Champion Problem}: despite achieving the highest baseline accuracy of 99.98\%, the Random Forest model suffers a catastrophic 73\% accuracy drop under minimal FGSM perturbation ($\epsilon=0.01$), whereas the CNN maintains 95.48\% accuracy, which is a 68.66 percentage-point robustness advantage. This contradicts conventional model selection strategies and shows that baseline accuracy can be actively misleading in security-critical applications. Second, reconnaissance attacks—including Ping Sweep, OS Scan, and Vulnerability Scan—exhibit universal vulnerability across all architectures, with CLEVER scores consistently below 1.5 compared to values above 8.0 for Brute Force classes. These observations underscore that vulnerability is not uniform across attack types and that class-specific mitigation strategies are necessary.

In this paper, we seek answers to the following research questions:

\begin{compactitem}
\item \textit{RQ1:} How does adversarial robustness differ across fundamentally distinct ML architectures used in network intrusion detection systems?

\item \textit{RQ2:} Is high baseline detection accuracy a reliable indicator of adversarial robustness in ML-based NIDS?

\item \textit{RQ3:} How does adversarial vulnerability vary across attack categories and individual intrusion classes in NIDS?
\end{compactitem}

Concretely, the contributions of this paper are 

\begin{compactitem}

\item We formally define and empirically validate the inverse relationship between baseline accuracy and adversarial robustness in NIDS. Our analysis demonstrates that conventional evaluation metrics systematically mislead model selection for security-critical deployments. 

\item We provide a systematic study linking architectural properties, including gradient accessibility, decision boundary geometry, representation learning capacity, and inherent regularization, to adversarial robustness outcomes. This establishes a principled foundation for robust NIDS architecture selection.

\end{compactitem}

We show that adversarial robustness varies sharply across attack categories, with reconnaissance attacks suffering up to 78\% F1-score degradation, compared to only 0.04\% for brute force attacks under similar perturbation conditions. This provides actionable guidance for prioritizing defensive hardening. We introduce a comprehensive evaluation methodology integrating adversarial attacks, certified robustness metrics, perturbability scoring, distribution shift analysis, and reconstruction-based assessment. This framework establishes a new standard for evaluating security-critical machine learning systems.

\section{RELATED WORK}
\label{sec:2RW}
In this section, we review prior research across four main areas: (1) machine learning approaches to intrusion detection, (2) adversarial machine learning, (3) adversarial attacks designed specifically for NIDS, and (4) robustness evaluation methodologies. Together, these threads highlight several important gaps that motivate our architectural analysis.

\subsection{Machine Learning for Intrusion Detection}

Machine learning has played a central role in intrusion detection for more than a decade, with early research focusing on classical algorithms such as decision trees, SVMs, and ensemble models \cite{ref1,ref20}, followed by strong success of hierarchical decision tree models such as Random Forests on benchmark datasets, helping establish them as a standard baseline \cite{ref1}.

Deep learning significantly reshaped the field of intrusion detection \cite{ref3}. Subsequent works explored convolutional and recurrent architectures to model spatial and temporal patterns in packets \cite{ref16,ref18}. These neural approaches consistently outperformed traditional models and became widely adopted in both academic research and commercial NIDS deployments.

Despite these advances, strong benchmark accuracy does not necessarily imply robustness in real-world settings. Sharafaldin \textit{et al.}~\cite{ref2} highlighted several dataset limitations that lead to overly optimistic performance estimates; models trained on outdated or synthetic traffic often fail when confronted with modern attacks. The ACI-IoT-2023 dataset addresses some of these issues by offering IoT traffic with realistic class imbalance \cite{ref22}. 

\subsection{Adversarial Machine Learning}

Adversarial attacks create very small, often imperceptible perturbations could cause deep neural networks to misclassify with high confidence \cite{ref23}. This discovery prompted extensive research into adversarial attack methods and defense strategies.

Goodfellow \textit{et al.}\cite{ref4} introduced the Fast Gradient Sign Method (FGSM), a computationally efficient single-step attack that perturbs inputs along the loss gradient direction. Although effective, FGSM often yields suboptimal adversarial examples. Madry \textit{et al.}~\cite{ref5} strengthened this line of work with Projected Gradient Descent (PGD), an iterative attack that remains the standard benchmark for robustness evaluation. They performed adversarial training which is augmenting training sets with adversarial examples. This way, models can substantially improved robustness. Other research explored decision-based and score-based attacks that rely only on model outputs rather than gradients \cite{ref6}. Papernot \textit{et al.}~\cite{ref23} showed that adversarial examples often transfer between models, enabling black-box attacks using surrogate networks. This transferability underscores a critical security concern that even models with hidden parameters are not necessarily protected.

\subsection{Adversarial Attacks on NIDS}

Although adversarial ML initially matured within computer vision, its relevance to intrusion detection quickly became clear \cite{ALHAJJAR2021115782}. Corona \textit{et al.}~\cite{ref10} provided one of the earliest taxonomies of adversarial attacks against IDS, distinguishing between evasion and poisoning strategies. Empirical studies soon verified these vulnerabilities through demonstrating that gradient-based attacks could reliably bypass deep-learning-based NIDS~\cite{ref11}.

Since adversarial perturbations must preserve packet validity and malicious behavior in NIDS adversarial modeling must incorporate realistic constraints, which limits the permissible perturbation space \cite{ref13}. Network traffic introduces structural constraints that differentiate NIDS attacks from those in the image domain. Feature-space perturbations may not correspond to feasible packet-level modifications \cite{ref14}. Nevertheless, practical evasion remains achievable under adversarial conditions and reported significant accuracy degradation across all tested models \cite{ref15}. What remains unclear in this literature is \emph{why} certain architectures fail more severely than others. While prior studies show that attacks succeed, they rarely examine architectural factors such as decision boundary geometry, gradient accessibility, or representation learning that fundamentally drive robustness differences.

\subsection{Robustness Evaluation Methods}
Evaluating adversarial robustness involves more than measuring accuracy under simple attacks. Many defenses collapse under stronger, adaptive attacks \cite{ref6}, revealing that robustness claims often reflect evaluation weaknesses rather than true resilience.

CLEVER score estimates the minimum adversarial perturbation required for misclassification using extreme value theory \cite{ref23}. It is attack-agnostic and therefore suitable for cross-model comparisons independent of specific attack algorithms. However, robustness to distribution shift is equally important. Quinonero-Candela \textit{et al.}\cite{ref22} formalized covariate shift and label shift as distinct challenges. These issues are especially relevant for NIDS, where network behavior evolves continuously and attack frequencies fluctuate. Such natural variations can degrade performance even without adversarial influence.


\subsection{Research Gap}
The existing body of research focuses on a few key aspects: (1) ML-based NIDS perform well on benchmark datasets; (2) adversarial attacks can significantly weaken these models; and (3) multiple metrics exist for quantifying robustness. However, several critical gaps remain.

Prior work has not systematically compared adversarial robustness across fundamentally different model families such as neural networks and tree-based ensembles using a consistent evaluation methodology to investigate the architectural weaknesses. Instead, studies typically focus on individual architectures or employ incompatible attack settings. Furthermore, the relationship between baseline accuracy and adversarial robustness remains poorly understood. The long-standing assumption that higher accuracy implies greater model quality does not hold in adversarial contexts, as our results demonstrate. Also, robustness analyses rarely examine differences across attack categories. Aggregate metrics often obscure substantial category-level variability, which can lead to misguided defensive priorities.

This paper fills these gaps through a unified, multi-dimensional robustness evaluation and a cross-architectural comparison of fundamentally distinct learning paradigms. Our analysis reveals the \textit{False Champion Problem} and proposes more principled guidelines for designing robust NIDS.

\section{METHODOLOGY}
\label{sec:3methodology}
\begin{figure}[t]
\centering
\includegraphics[width=\columnwidth]{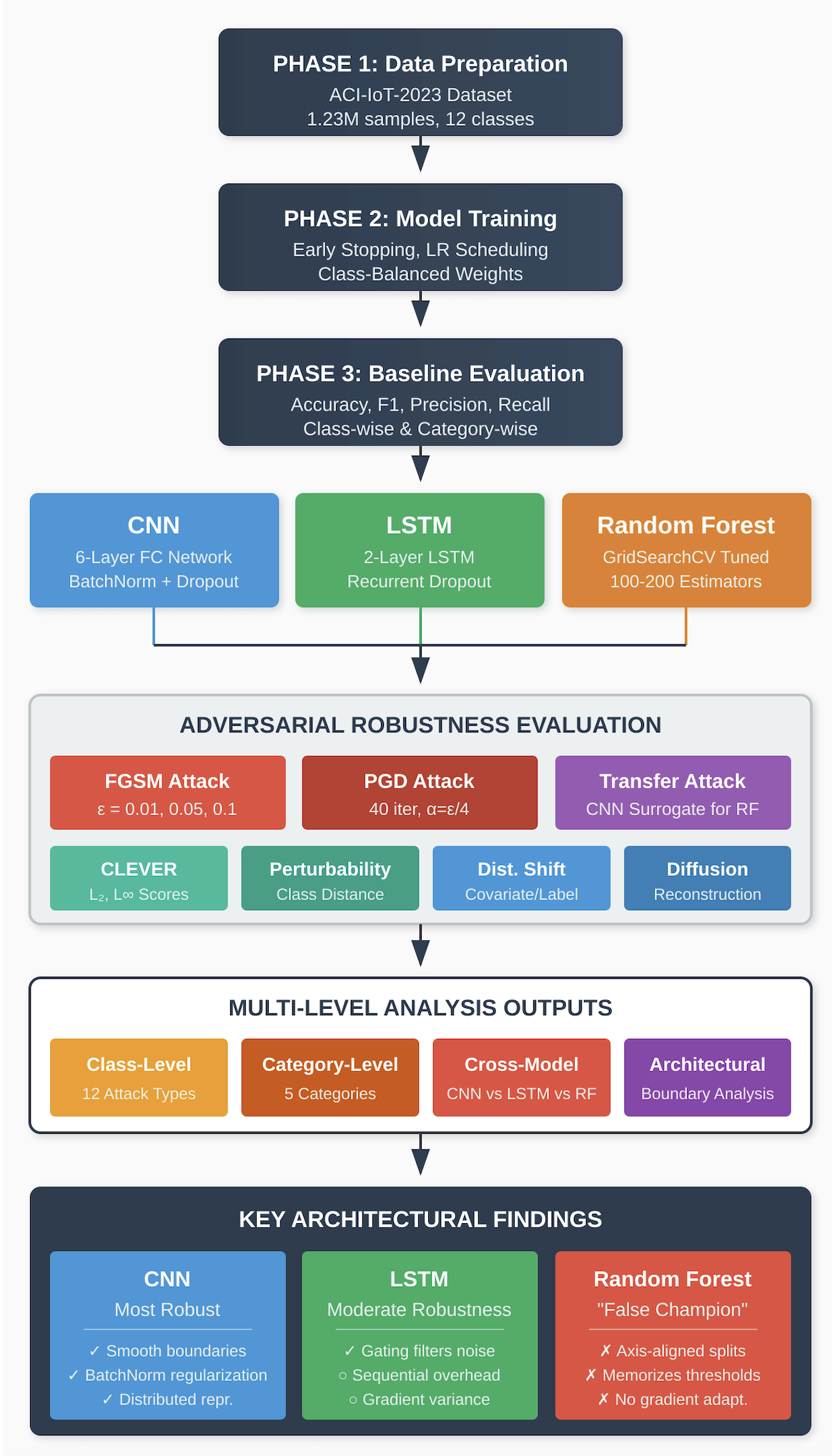}
\caption{Adversarial Robustness Evaluation Framework}
\label{fig:framework}
\end{figure}

In this section, we outline how our multi-dimensional adversarial robustness evaluation framework works for machine learning-based Network Intrusion Detection Systems. We first walk through the dataset we used, explain our feature engineering approach, describe each model architecture, detail our adversarial attack implementations, and lay out the full robustness assessment process. Fig.~\ref{fig:framework} shows the complete pipeline, covering data preparation through model training, baseline evaluation, adversarial testing, and the final comparative analysis.

\subsection{Dataset Description}

ACI-IoT-2023 dataset \cite{ref22} was built specifically for intrusion detection research in IoT and smart environment settings. Table~\ref{tab:dataset} back in Section~I breaks down the full dataset composition. The data contains 1,231,411 labeled network flow samples across 12 different attack classes, which we grouped into five broader categories: Benign, Reconnaissance (Recon), Denial-of-Service (DoS), Brute Force, and Spoofing. Having this two-level structure lets us evaluate vulnerabilities at both a fine-grained class level and a more general category level. This perspective is often lost when studies display only aggregate numbers.

ACI-IoT-2023 dataset is imbalanced but reflects real network traffic with attacks performed on IoT systems. Port Scan makes up the largest chunk at 35.8\% (441,349 samples), while ARP Spoofing barely registers at less than 0.01\% (just 6 samples). The other classes fall somewhere in between: Benign traffic sits at 26.9\% (331,496 samples), ICMP Flood at 12.6\% (154,981), and Dictionary Attack at 3.0\% (37,089). This kind of skewed distribution is useful for our purposes since it reflects what security systems face in the real world, where some attacks show up constantly while others are rare. It also lets us explore whether adversarial vulnerability has anything to do with how common a class is a question that matters quite a bit when deciding where to focus defensive resources.

The dataset covers a good range of modern IoT attack behaviors such as reconnaissance probing that often comes before bigger attacks, DoS floods that hammer protocol weaknesses, brute force credential stuffing, and spoofing attacks that meddles with network identities. This variety means our robustness findings should hold up across various kinds of threats IoT deployments face today.

\subsection{Feature Engineering and Preprocessing}

We extracted 30 features that capture network flow behavior from several different angles. Table~\ref{tab:features} lists these features by category and explains why each one matters for detecting intrusions.

\begin{table}[!t]
\centering
\caption{Feature Categories and Security-Relevant Descriptions}
\label{tab:features}
\begin{tabular}{p{1.8cm}p{6cm}}
\hline
\textbf{Category} & \textbf{Features and Description} \\
\hline
Flow Identifiers & Flow ID, Source Port, Destination Port, Protocol, Timestamp: these pin down individual flows and let us do protocol-specific analysis \\
\hline
Duration & Flow Duration: helps tell apart quick burst attacks from longer-running connections \\
\hline
Packet Volume & Total Forward Packets, Total Backward Packets: lopsided counts often point to scanning or flooding \\
\hline
Packet Length Statistics & Forward/Backward Packet Length (Max, Min, Mean, Std): the payload patterns here can reveal attack signatures \\
\hline
Throughput & Flow Bytes/s, Flow Packets/s: rate information is key for catching DoS attacks \\
\hline
Timing Patterns & Flow IAT Mean, Std, Max, Min: the timing between packets looks very different for automated attacks versus normal human traffic \\
\hline
Activity Cycles & Active/Idle Mean, Std, Max, Min: how connections persist tells us about what stage an attack is in \\
\hline
\end{tabular}
\end{table}

The preprocessing pipeline works as follows.

\subsubsection{Categorical Encoding}
We converted categorical features like Flow ID and Protocol to numbers using label encoding. For Protocol specifically, we encoded the transport layer types (TCP, UDP, ICMP) as integers, which keeps their discriminative power for spotting protocol-specific attacks.

\subsubsection{Temporal Transformation}
We turned timestamps into Unix epoch format (seconds since January 1, 1970). This preserves the time ordering while making it easy to do math operations like computing flow durations and analyzing temporal patterns.

\subsubsection{Missing Value Treatment}
Division operations sometimes produce infinite values (like when Flow Bytes/s divides by a near-zero duration). We flagged these as NaN and dropped them. Taking this conservative route avoids numerical problems during gradient calculations without introducing fake values through imputation.

\subsubsection{Feature Standardization}
We standardized all numerical features using z-scores:
\begin{equation}
x_{norm} = \frac{x - \mu}{\sigma}
\label{eq:zscore}
\end{equation}
where $\mu$ and $\sigma$ are the mean and standard deviation calculated only from training data. This accomplishes three things: it keeps gradient magnitudes consistent during neural network training, makes perturbation effects comparable across features when generating adversarial examples, and stops features with naturally larger scales from overpowering the learning process.

\subsubsection{Data Partitioning}
We split the data 80/20 for training and testing, using stratified sampling on the labels to keep class proportions the same in both sets. This gave us 985,128 training samples and 246,283 test samples.

\subsubsection{Class Imbalance Mitigation} The severe class imbalance needed addressing during training, so we used 40\% minority class repopulation through random oversampling with replacement. Basically, we identified all the non-majority classes, resampled them to make up 40\% of an augmented training set, and combined that with the original training data. We only did this to training data; the test set stayed untouched to keep evaluation fair. On top of that, we used class-balanced loss weighting that gives each class a weight inversely proportional to how often it appears:
\begin{equation}
w_c = \frac{N}{C \cdot n_c}
\label{eq:classweight}
\end{equation}
where $N$ is the total training sample count, $C$ is the number of classes, and $n_c$ is how many samples belong to class $c$.

\subsection{Model Architectures}

We tested three classifiers that work in fundamentally different ways: two gradient-based deep neural networks (CNN and LSTM) and one gradient-free ensemble method (Random Forest). Picking architectures this different from each other lets us systematically investigate how model design affects adversarial robustness. Table~\ref{tab:architectures} summarizes the key settings for each.

\begin{table}[!t]
\centering
\caption{Model Architecture and Training Hyperparameters}
\label{tab:architectures}
\begin{tabular}{lccc}
\hline
\textbf{Parameter} & \textbf{CNN} & \textbf{LSTM} & \textbf{RF} \\
\hline
\multicolumn{4}{l}{\textit{Architecture Parameters}} \\
Hidden Layers & 6 FC & 2 LSTM + 3 FC & --- \\
Hidden Dimension & 256 & 256 & --- \\
Activation Function & ReLU & ReLU/Tanh & --- \\
Dropout Rate & 0.3 & 0.2 (recurrent) & --- \\
Batch Normalization & Yes & No & --- \\
Number of Estimators & --- & --- & 100--200 \\
Maximum Depth & --- & --- & None/20/30 \\
Min Samples Split & --- & --- & 2/5 \\
\hline
\multicolumn{4}{l}{\textit{Training Parameters}} \\
Optimizer & Adam & Adam & --- \\
Initial Learning Rate & $5 \times 10^{-4}$ & $10^{-3}$ & --- \\
Batch Size & 512 & 512 & --- \\
Maximum Epochs & 50 & 50 & --- \\
Early Stopping Patience & 10 & 10 & --- \\
LR Scheduler Patience & 5 & 5 & --- \\
LR Reduction Factor & 0.5 & 0.5 & --- \\
Minimum Learning Rate & $10^{-6}$ & $10^{-6}$ & --- \\
Gradient Clip Norm & 1.0 & 1.0 & --- \\
Class Weighting & Balanced & Balanced & Balanced \\
\hline
\end{tabular}
\end{table}

\subsubsection{Convolutional Neural Network (CNN)}

Our CNN uses a deep fully-connected architecture with convolutional-style regularization, tailored for tabular network traffic data. We stack six fully-connected layers, each with 256 hidden units, building up a deep representation learning pipeline. Each hidden layer $l$ computes:
\begin{equation}
h^{(l)} = \text{Dropout}\left(\text{ReLU}\left(\text{BN}\left(W^{(l)} h^{(l-1)} + b^{(l)}\right)\right)\right)
\label{eq:cnn_layer}
\end{equation}
where $W^{(l)}$ and $b^{(l)}$ are the learnable weights and biases, BN stands for Batch Normalization, and the network takes in the 30-dimensional feature vector as $h^{(0)} = x$.

Batch Normalization \cite{ref23} shifts layer activations to have zero mean and unit variance, then rescales them with learnable parameters $\gamma$ (scale) and $\beta$ (shift):
\begin{equation}
\text{BN}(z) = \gamma \cdot \frac{z - \mathbb{E}[z]}{\sqrt{\text{Var}[z] + \epsilon}} + \beta
\label{eq:batchnorm}
\end{equation}
This helps in several ways: training becomes more stable because internal covariate shift goes down, we can use higher learning rates, and it acts as a regularizer to reduce overfitting. For adversarial robustness specifically, Batch Normalization tends to smooth out the loss landscape, which might create gentler decision boundaries that need bigger perturbations to cross.

\textbf{Dropout} at a rate of 0.3 randomly zeros out 30\% of activations during training. This stops neurons from co-adapting and pushes the network toward learning distributed representations. By forcing redundant feature detectors to form, dropout may boost robustness since the model can't lean too heavily on any single feature pathway that an attacker could exploit.

The final layer maps the 256-dimensional representation down to 12 classes without any activation:
\begin{equation}
\hat{y} = W^{(out)} h^{(L)} + b^{(out)}
\label{eq:output}
\end{equation}
These raw logits $\hat{y}$ then go into the softmax cross-entropy loss.

\noindent \textbf{Training Procedure}: We used Adam \cite{ref23} as our optimizer with a starting learning rate of $5 \times 10^{-4}$. We went lower than the usual default because deeper architectures need more careful convergence. The loss function pairs softmax cross-entropy with class-balanced weighting:
\begin{equation}
\mathcal{L} = -\sum_{i=1}^{N} w_{y_i} \log \frac{\exp(\hat{y}_i[y_i])}{\sum_{c=1}^{C} \exp(\hat{y}_i[c])}
\label{eq:loss}
\end{equation}
where $w_{y_i}$ weights the loss according to sample $i$'s true label $y_i$.

We clipped gradients at max norm 1.0 to prevent explosion in our deep network. The learning rate drops by half whenever validation accuracy plateaus for 5 epochs, down to a floor of $10^{-6}$. Training stops early if validation accuracy doesn't improve for 10 epochs, and we restore whichever checkpoint performed best.

\subsubsection{Long Short-Term Memory Network (LSTM)}

The LSTM architecture \cite{ref18} treats network flows as sequential data, feeding the 30 features as a single-timestep sequence. While this isn't the typical way LSTMs get used, it lets the gating mechanisms work their magic on feature interactions through selective filtering.

We stacked two LSTM layers with 256 hidden units each. At every timestep, the LSTM cell runs through these computations:
\begin{align}
f_t &= \sigma(W_f \cdot [h_{t-1}, x_t] + b_f) \label{eq:forget} \\
i_t &= \sigma(W_i \cdot [h_{t-1}, x_t] + b_i) \label{eq:input} \\
\tilde{C}_t &= \tanh(W_C \cdot [h_{t-1}, x_t] + b_C) \label{eq:candidate} \\
C_t &= f_t \odot C_{t-1} + i_t \odot \tilde{C}_t \label{eq:cell} \\
o_t &= \sigma(W_o \cdot [h_{t-1}, x_t] + b_o) \label{eq:output_gate} \\
h_t &= o_t \odot \tanh(C_t) \label{eq:hidden}
\end{align}
Here $\sigma$ is the sigmoid function, $\odot$ means element-wise multiplication, $f_t$ is the forget gate, $i_t$ is the input gate, $o_t$ is the output gate, $C_t$ is the cell state, and $h_t$ is the hidden state.

What makes this interesting for robustness is the built-in noise filtering. The forget gate ($f_t$) decides what to throw away from the cell state, while the input gate ($i_t$) controls what new information gets in. Our hypothesis is that this selective filtering might help against adversarial attacks by treating small perturbations as noise to filter out rather than meaningful signal to pass along.

Between LSTM layers, we applied recurrent dropout at 0.2, dropping the same units across timesteps so we don't mess up temporal dynamics. The final hidden state $h_T$ goes through three fully-connected layers (256$\rightarrow$256$\rightarrow$12) with ReLU activations for the final classification.

Training worked the same way as for the CNN, except we used a learning rate of $10^{-3}$ (which is more standard for LSTMs), along with class-balanced weighting, gradient clipping, learning rate scheduling, and early stopping.

\subsubsection{Random Forest (RF)}

Random Forest \cite{ref20} takes a completely different approach from the gradient-based neural networks. It builds an ensemble of $T$ decision trees, each trained on a bootstrap sample of the data using random feature subsets at each split. Predictions come from majority voting across all trees:
\begin{equation}
\hat{y} = \text{mode}\left\{h_t(x) : t = 1, \ldots, T\right\}
\label{eq:rf_predict}
\end{equation}
where $h_t(x)$ is what tree $t$ predicts for input $x$.

We ran GridSearchCV to find good hyperparameters, searching over:
\begin{itemize}
\item Number of estimators: $T \in \{100, 200\}$
\item Maximum tree depth: $d_{max} \in \{\text{None}, 20, 30\}$
\item Minimum samples per split: $n_{split} \in \{2, 5\}$
\end{itemize}
That's 12 combinations total, each evaluated with 3-fold stratified cross-validation. We picked whichever configuration got the highest weighted F1 score. 
For class imbalance, we weight samples inversely to class frequency during tree building.

\textbf{Decision Boundary Geometry}: Random Forests carve up the feature space using axis-aligned cuts. Each internal node picks the feature $j$ and threshold $\tau$ that maximize information gain:
\begin{equation}
(j^*, \tau^*) = \arg\max_{j, \tau} \left[ H(S) - \frac{|S_L|}{|S|} H(S_L) - \frac{|S_R|}{|S|} H(S_R) \right]
\label{eq:split}
\end{equation}
where $H(\cdot)$ is entropy, $S$ is the current node's sample set, and $S_L$, $S_R$ are what goes left and right after the split.

This creates decision surfaces made up of axis-aligned hyperplanes, which behave very differently from what neural networks produce:

\textit{Neural Networks}: Their decision boundaries are smooth, continuous hypersurfaces that emerge from stacking nonlinear transformations. Crossing from one class region to another means traversing a gradient field, and how much perturbation you need depends on the boundary's local curvature.

\textit{Random Forests}: Their decision boundaries are choppy, axis-aligned step functions. A tiny tweak to just one feature can push a sample past a threshold, instantly changing how a tree routes it and potentially flipping the ensemble's prediction. These discrete, threshold-based boundaries create weak points.

This difference is central to what we're testing: we think Random Forests' threshold-based decisions, while great for efficient partitioning and interpretability, create fragile decision boundaries that attackers can exploit with small, targeted perturbations.

\subsection{Adversarial Attack Methods}

In our \textbf{threat model}, we adopt a white-box evasion attack model for neural network evaluation, where the adversary has complete knowledge of the target model's architecture, parameters, and gradients. This represents the strongest adversarial capability and provides an upper bound on vulnerability—models that withstand white-box attacks will necessarily resist weaker black-box variants. For Random Forest, we relax this to a transfer-based gray-box setting where the attacker trains a surrogate model on similar data but lacks direct access to the target classifier.

Our attacks operate in normalized feature space rather than raw network traffic. This is a standard evaluation protocol in adversarial ML research that measures model-level robustness independent of domain-specific packet manipulation constraints. While this approach does not guarantee that every perturbation maps to a physically realizable network flow modification, the perturbation magnitudes we test ($\epsilon \leq 0.1$ in normalized space) correspond to modest changes in flow statistics, which are adjustments to packet timing, sizes, or counts that an attacker could plausibly implement while maintaining attack functionality. We acknowledge that a complete real-world evasion pipeline would require additional domain constraints, which we identify as a direction for future work.

\subsubsection{Fast Gradient Sign Method (FGSM)}

FGSM \cite{ref4} creates adversarial examples in one shot by perturbing along the gradient direction:
\begin{equation}
x_{adv} = x + \epsilon \cdot \text{sign}\left(\nabla_x \mathcal{L}(\theta, x, y)\right)
\label{eq:fgsm}
\end{equation}
where $x$ is the original input, $\epsilon$ sets how big the perturbation can be, $\mathcal{L}$ is the cross-entropy loss, $\theta$ represents model parameters, and $y$ is the true label. Taking the sign of the gradient gives us unit $L_\infty$ norm in each dimension, which maximizes how much we increase the loss for a given perturbation budget.

We tested three perturbation levels covering a range of attack intensities:
\begin{itemize}
\item $\epsilon = 0.01$: Just 1\% of the normalized feature range. These minimal changes might not even be noticeable in raw feature space.
\item $\epsilon = 0.05$: A 5\% perturbation. Moderate attack strength.
\item $\epsilon = 0.1$: A 10\% perturbation. Aggressive attacks that significantly alter flow characteristics.
\end{itemize}

\subsubsection{Projected Gradient Descent (PGD)}

PGD \cite{ref5} builds on FGSM by iterating multiple steps, essentially solving the inner maximization problem from adversarial training:
\begin{equation}
x_{adv}^{t+1} = \Pi_{x + \mathcal{S}} \left( x_{adv}^{t} + \alpha \cdot \text{sign}\left(\nabla_x \mathcal{L}(\theta, x_{adv}^{t}, y)\right) \right)
\label{eq:pgd}
\end{equation}
where $\Pi_{x + \mathcal{S}}$ projects everything back into the $L_\infty$ $\epsilon$-ball $\mathcal{S}$ around the original input $x$, making sure $\|x_{adv} - x\|_\infty \leq \epsilon$ at every step. The step size $\alpha$ controls how granular each update is.

Our PGD setup used:
\begin{itemize}
\item $\alpha = \epsilon / 4$, scaling with the perturbation budget for adaptive step size
\item performed $T = 40$ enough steps for the optimization to converge
\item Started from the clean input for deterministic evaluation
\item Clipped to $[x - \epsilon, x + \epsilon]$ after every step
\end{itemize}

PGD is considerably stronger than FGSM since it iteratively refines perturbations toward locally optimal adversarial examples. The projection keeps perturbations in bounds while letting the optimization explore the full boundary of the $\epsilon$-ball. Most researchers consider PGD the gold standard for empirical robustness testing. Usually, if a defense holds up against PGD, it usually generalizes to other $L_\infty$-bounded attacks~\cite{ref5}.

\subsubsection{Transfer Attacks for Random Forest}

Random Forests do not have a differentiable loss function, so we cannot compute gradients directly. The ensemble voting and threshold-based tree routing just do not give us anything to work with for optimization-based attacks. Instead, we leveraged the well-known fact that adversarial examples transfer across different model architectures \cite{ref19}.

Specifically, we generated adversarial examples against the CNN using both FGSM and PGD, then fed those same perturbed inputs to the Random Forest:
\begin{equation}
x_{adv}^{RF} = \text{Attack}_{CNN}(x, y; \epsilon)
\label{eq:transfer}
\end{equation}
where $\text{Attack}_{CNN}$ means running FGSM or PGD against the CNN surrogate.

This approach reflects realistic black-box attack scenarios:
\begin{enumerate}
\item The attacker cannot peek at the target Random Forest's structure, feature importance, or decision paths.
\item Attackers can train substitute models on similar data, either by collecting their own data or through model extraction attacks.
\item Adversarial perturbations often work across architectures because models trained on similar data tend to carve up the input space in similar ways.
\end{enumerate}

We used the CNN as our surrogate training model because it is differentiable, performs well enough to give meaningful gradients, and is architecturally very different from Random Forests. This tests cross-paradigm transferability rather than transfer within the same model family.

\subsection{Robustness Evaluation Framework}

Our evaluation goes well beyond just measuring accuracy drops. We built a multi-dimensional framework with six components, each shedding light on different aspects of vulnerability. Algorithm~\ref{alg:evaluation} lays out the full procedure.

\begin{algorithm}[!t]
\caption{Multi-Dimensional Adversarial Robustness Evaluation Framework}
\label{alg:evaluation}
\begin{algorithmic}[1]
\REQUIRE Trained models $\mathcal{M} = \{M_{\mathrm{CNN}}, M_{\mathrm{LSTM}}, M_{\mathrm{RF}}\}$
\REQUIRE Test data $(X_{\mathrm{test}}, y_{\mathrm{test}})$, perturbation budgets $\mathcal{E} = \{0.01, 0.05, 0.1\}$
\ENSURE Comprehensive robustness metrics for each model

\FOR{each model $M \in \mathcal{M}$}
    \STATE \textbf{Phase 1: Baseline Performance}
    \STATE $\hat{y} \leftarrow M(X_{\mathrm{test}})$
    \STATE Compute Accuracy, Precision, Recall, F1 (aggregate)
    \STATE Compute class-wise metrics for all 12 classes
    \STATE Compute category-wise metrics for 5 categories
    
    \STATE \textbf{Phase 2: Adversarial Attack Evaluation}
    \FOR{each $\epsilon \in \mathcal{E}$}
        \STATE $X_{\mathrm{fgsm}} \leftarrow \text{FGSM}(M, X_{\mathrm{test}}, y_{\mathrm{test}}, \epsilon)$  
        \STATE $X_{\mathrm{pgd}} \leftarrow \text{PGD}(M, X_{\mathrm{test}}, y_{\mathrm{test}}, \epsilon, \alpha=\epsilon/4, T=40)$  
        \STATE Evaluate metrics on $X_{\mathrm{fgsm}}$ and $X_{\mathrm{pgd}}$
        \STATE Compute class-wise and category-wise degradation
    \ENDFOR
    
    \STATE \textbf{Phase 3: Certified Robustness Metrics}
    \FOR{$c = 1$ \TO $12$}
        \STATE Sample $N = 100$ instances from class $c$
        \STATE $\mathrm{CLEVER}_c \leftarrow$ mean pairwise $L_2$ distance
        \STATE $\mathrm{Perturb}_c \leftarrow$ intra-class dispersion score
    \ENDFOR
    
    \STATE \textbf{Phase 4: Distribution Shift Robustness}
    \FOR{each $\sigma \in \{0.01, 0.05, 0.1\}$}
        \STATE $X_{\mathrm{cov}} \leftarrow X_{\mathrm{test}} + \mathcal{N}(0, \sigma^2 I)$   
        \STATE Evaluate metrics on $X_{\mathrm{cov}}$
    \ENDFOR
    \STATE $X_{\mathrm{label}} \leftarrow$ oversample minority classes $10 \times$
    \STATE Evaluate metrics on $X_{\mathrm{label}}$   
    
    \STATE \textbf{Phase 5: Generalization Assessment}
    \FOR{each class $c$}
        \STATE $\mu_c \leftarrow$ class centroid
        \STATE Generate 200 samples: $x_{\mathrm{syn}} \sim \mathcal{N}(\mu_c, 0.1^2 I)$
        \STATE Evaluate classification accuracy on synthetic samples
    \ENDFOR
\ENDFOR

\STATE \textbf{Phase 6: Cross-Model Comparative Analysis}
\STATE Generate comparative visualizations across all models
\STATE Compute degradation heatmaps, radar charts, and trend analyses
\STATE Perform statistical comparison of robustness profiles

\end{algorithmic}
\end{algorithm}

\subsubsection{Baseline Performance Metrics}

Standard classification metrics give us a baseline to measure adversarial degradation against. From the confusion matrix with true positives ($TP$), false positives ($FP$), true negatives ($TN$), and false negatives ($FN$), we calculate:
\begin{align}
\text{Accuracy} &= \frac{TP + TN}{TP + TN + FP + FN} \label{eq:accuracy} \\
\text{Precision}_c &= \frac{TP_c}{TP_c + FP_c} \label{eq:precision} \\
\text{Recall}_c &= \frac{TP_c}{TP_c + FN_c} \label{eq:recall} \\
\text{F1}_c &= 2 \cdot \frac{\text{Precision}_c \cdot \text{Recall}_c}{\text{Precision}_c + \text{Recall}_c} \label{eq:f1}
\end{align}

We compute these at three levels of granularity:
\begin{itemize}
\item \textbf{Aggregate}: Weighted averages across all classes for an overall performance picture
\item \textbf{Class-wise}: Separate metrics for each of the 12 attack types, showing detailed detection capabilities
\item \textbf{Category-wise}: Metrics for each of the 5 broader categories, which is easier to interpret operationally
\end{itemize}

Looking at multiple levels matters because aggregate stats can hide problems. A model hitting 95\% overall accuracy might completely fail on specific attack classes that security teams really care about.

\subsubsection{Adversarial Attack Evaluation}

For every combination of model, attack type (FGSM, PGD), and perturbation size ($\epsilon \in \{0.01, 0.05, 0.1\}$), we track:
\begin{compactitem}
\item \textbf{Accuracy degradation}: $\Delta_{acc} = \text{Acc}_{baseline} - \text{Acc}_{adversarial}$
\item \textbf{F1 degradation}: $\Delta_{F1} = \text{F1}_{baseline} - \text{F1}_{adversarial}$
\item \textbf{Class-wise vulnerability}: How each class's metrics change, pinpointing the most susceptible attack types
\item \textbf{Category-wise patterns}: Vulnerability aggregated by attack category
\end{compactitem}

Comparing baseline to adversarial performance tells us about robustness level of the model. Small drops mean genuine resilience, while big drops signal architectural weaknesses. How degradation scales across $\epsilon$ values is informative too gradual decline suggests smooth decision boundaries, while sudden collapse points to brittle thresholds.

\subsubsection{CLEVER Score Computation}

The Cross-Lipschitz Extreme Value for nEtwork Robustness (CLEVER) metric \cite{ref23} gives us robustness estimates that don't depend on any particular attack algorithm. Unlike empirical robustness (how accuracy holds up against specific attacks), CLEVER characterizes intrinsic vulnerability by approximating the minimum perturbation needed to cause misclassification.

For each class $c$, we:
\begin{enumerate}
\item Take $N = 100$ samples belonging to class $c$ from the test set
\item Calculate all pairwise $L_2$ distances between these samples
\item Report the average distance as the CLEVER score:
\begin{equation}
\text{CLEVER}_c = \frac{1}{N(N-1)} \sum_{i \neq j} \|x_i - x_j\|_2
\label{eq:clever}
\end{equation}
\end{enumerate}

Lower CLEVER scores mean the class's samples cluster tightly in feature space, putting them closer to decision boundaries and making them easier targets. Higher scores suggest more room before perturbations could push samples across class boundaries.

\subsubsection{Perturbability Analysis}

While CLEVER focuses on proximity to boundaries, perturbability scores look at how spread out each class is in feature space:
\begin{equation}
\text{Perturbability}(c) = \frac{1}{|C_c|^2} \sum_{x_i, x_j \in C_c} \|x_i - x_j\|_2
\label{eq:perturb}
\end{equation}
where $C_c$ is all test samples from class $c$.

Classes with high perturbability take up more feature space, which might mean they can absorb more perturbation before samples get pushed across boundaries. Putting CLEVER and perturbability together gives a better picture:
\begin{compactitem}
\item High CLEVER + High Perturbability presents a robust class, sitting far from boundaries with room to spare
\item Low CLEVER + Low Perturbability presents a vulnerable class, tightly clustered near boundaries
\item Mixed patterns for both present more complicated situations that need careful interpretation
\end{compactitem}

\subsubsection{Distribution Shift Robustness}

Real deployments expose models to natural data variations that are different from adversarial attacks. We tested robustness against two types of shift that commonly happen in operational NIDS settings:

\noindent \textbf{a. Covariate Shift}: Adding Gaussian noise simulates things like sensor drift, measurement noise, and changing network conditions. The noise changes feature distributions while keeping label meanings intact:
\begin{equation}
x_{shifted} = x + \eta, \quad \eta \sim \mathcal{N}(0, \sigma^2 I)
\label{eq:covariate}
\end{equation}
We tested $\sigma \in \{0.01, 0.05, 0.1\}$, covering mild to substantial perturbation. Models that handle covariate shift well should keep working as network conditions evolve.

\noindent \textbf{b. Label Shift}: We oversampled minority classes (ARP Spoofing, Dictionary Attack, UDP Flood) by 10$\times$ to simulate what happens when attack prevalence changes between training and deployment. This checks whether models stay calibrated and maintain per-class accuracy when $P(Y)$ shifts but $P(X|Y)$ stays the same.

\subsubsection{Diffusion-Based Reconstruction}

To test generalization, we checked whether models actually learned robust class representations or just memorized training examples. For each class $c$:
\begin{enumerate}
\item We computed the class centroid: $\mu_c = \frac{1}{|C_c|} \sum_{x \in C_c} x$

\item Generated 200 synthetic samples by adding Gaussian noise around the centroid:
\begin{equation}
x_{syn} \sim \mathcal{N}(\mu_c, 0.1^2 I)
\label{eq:diffusion}
\end{equation}
\item Evaluated how accurately the model classified these synthetic samples
\end{enumerate}

High accuracy on reconstructed samples means the model learned generalizable representations that extend beyond specific training instances. Low accuracy suggests overfitting to particular feature patterns, which would make the model vulnerable to new attack variants within the same class.

\subsection{Experimental Configuration}

We ran all experiments on standardized hardware with fixed random seeds to make the work reproducible. NumPy and PyTorch seeds were both set to 42, and we enabled deterministic CUDA operations where possible.

\subsubsection{Neural Network Training}
We trained for up to 50 epochs with early stopping (patience 10), using batch size 512 and the hyperparameters from Table~\ref{tab:architectures}. Validation accuracy guided both early stopping and learning rate scheduling. After training, we loaded back whichever checkpoint had the best validation accuracy for final evaluation.

\subsubsection{Random Forest Training}
GridSearchCV worked through all 12 hyperparameter combinations (2 estimator counts $\times$ 3 max depths $\times$ 2 min samples) using 3-fold stratified cross-validation. We picked the configuration with the highest weighted F1 score.

\subsubsection{Evaluation Protocol}
Every metric comes from the held-out test partition (246,283 samples, 20\% of the data). We generated adversarial examples fresh for each attack configuration to make sure there's no information leakage between evaluation runs. Class-wise and category-wise metrics used the same test partition with appropriate filtering.

\subsubsection{Implementation}
The whole pipeline uses PyTorch 2.0 for neural networks and adversarial attack generation, scikit-learn 1.3 for Random Forest training and metric computation, plus the standard scientific Python stack (NumPy 1.24, Pandas 2.0, Matplotlib 3.7) for data processing and visualization. We saved all intermediate results such as confusion matrices, classification reports, adversarial accuracies, CLEVER scores, perturbability metrics for reproducibility checks. The full codebase, trained model checkpoints, and evaluation outputs are available for independent verification.

\section{EXPERIMENTAL RESULTS \& ANALYSIS}
\label{sec:experimental analysis & results}
This section presents our experimental findings from the multi-dimensional adversarial robustness evaluation. We start with baseline performance to establish each model's detection capabilities, then dive into how they hold up under FGSM and PGD attacks at various perturbation strengths. The results reveal some striking differences between neural network and ensemble approaches, which are differences that challenge conventional assumptions about model robustness.

The experimental results presented in this section address the three research questions introduced in the introduction section. Specifically, Sections~\ref{sub:E-A} and~\ref{sub:E-B}, primarily address RQ1, examining how adversarial robustness differs across ML architectures. Section~\ref{sub:E-B} further addresses RQ2 by contrasting baseline accuracy with adversarial degradation to analyze the False Champion Problem. Finally, Sections~\ref{sub:E-B} and~\ref{sub:E-C} address RQ3 through class-wise and category-wise vulnerability analysis.

\begin{table*}[!t]
\centering
\caption{Comprehensive Adversarial Robustness Evaluation Results Across All Attack Configurations}
\label{tab:comprehensive_results}
\renewcommand{\arraystretch}{1.2}
\small
\begin{tabular}{ll|cccc|cccc|cccc}
\toprule
\multirow{2}{*}{\textbf{Attack}} & \multirow{2}{*}{\textbf{$\epsilon$}} & \multicolumn{4}{c|}{\textbf{CNN}} & \multicolumn{4}{c|}{\textbf{LSTM}} & \multicolumn{4}{c}{\textbf{Random Forest}} \\
\cmidrule(lr){3-6} \cmidrule(lr){7-10} \cmidrule(lr){11-14}
 & & Acc & Prec & Rec & F1 & Acc & Prec & Rec & F1 & Acc & Prec & Rec & F1 \\
\midrule
\multicolumn{2}{l|}{\textit{Baseline (No Attack)}} & 0.991 & 0.993 & 0.991 & 0.992 & 0.994 & 0.995 & 0.994 & 0.994 & \textbf{1.000} & \textbf{1.000} & \textbf{1.000} & \textbf{1.000} \\
\midrule
\multirow{3}{*}{FGSM} 
 & 0.01 & \textbf{0.955} & \textbf{0.969} & \textbf{0.955} & \textbf{0.960} & 0.850 & 0.922 & 0.850 & 0.872 & 0.268 & 0.728 & 0.268 & 0.116 \\
 & 0.05 & \textbf{0.561} & \textbf{0.662} & \textbf{0.561} & \textbf{0.593} & 0.583 & 0.726 & 0.583 & 0.569 & 0.266 & 0.262 & 0.266 & 0.112 \\
 & 0.1 & 0.495 & \textbf{0.586} & 0.495 & \textbf{0.487} & \textbf{0.513} & 0.571 & \textbf{0.513} & 0.481 & 0.266 & 0.446 & 0.266 & 0.112 \\
\midrule
\multirow{3}{*}{PGD} 
 & 0.01 & \textbf{0.854} & \textbf{0.956} & \textbf{0.854} & \textbf{0.887} & 0.686 & 0.887 & 0.686 & 0.731 & 0.269 & 0.790 & 0.269 & 0.118 \\
 & 0.05 & 0.453 & \textbf{0.621} & 0.453 & 0.475 & \textbf{0.490} & 0.554 & \textbf{0.490} & \textbf{0.483} & 0.266 & 0.489 & 0.266 & 0.112 \\
 & 0.1 & 0.287 & \textbf{0.380} & 0.287 & \textbf{0.286} & \textbf{0.296} & 0.318 & \textbf{0.296} & 0.250 & 0.280 & 0.797 & 0.280 & 0.139 \\
\midrule
\multicolumn{14}{l}{\textit{Performance Degradation from Baseline (percentage points)}} \\
\midrule
FGSM & 0.1 & $-$49.6 & $-$40.7 & $-$49.6 & $-$50.5 & $-$48.1 & $-$42.4 & $-$48.1 & $-$51.3 & \cellcolor{red!25}\textbf{$-$73.4} & \cellcolor{red!25}\textbf{$-$55.4} & \cellcolor{red!25}\textbf{$-$73.4} & \cellcolor{red!25}\textbf{$-$88.8} \\
PGD & 0.1 & $-$70.4 & $-$61.3 & $-$70.4 & $-$70.6 & $-$69.8 & $-$67.7 & $-$69.8 & $-$74.4 & \cellcolor{red!25}\textbf{$-$72.0} & $-$20.3 & \cellcolor{red!25}\textbf{$-$72.0} & \cellcolor{red!25}\textbf{$-$86.1} \\
\bottomrule
\end{tabular}
\vspace{2mm}
\begin{flushleft}
\footnotesize
\textit{Note:} Bold values indicate best performance for each attack configuration. Red-highlighted cells show Random Forest's catastrophic degradation despite achieving highest baseline performance. CNN demonstrates most consistent robustness, retaining 96\% accuracy at $\epsilon=0.01$ under FGSM while Random Forest drops to 27\%. All metrics computed on held-out test set (246,283 samples).
\end{flushleft}
\end{table*}

\subsection{Baseline Performance Analysis}
\label{sub:E-A}

\begin{figure}[t]
\centering
\includegraphics[width=\columnwidth]{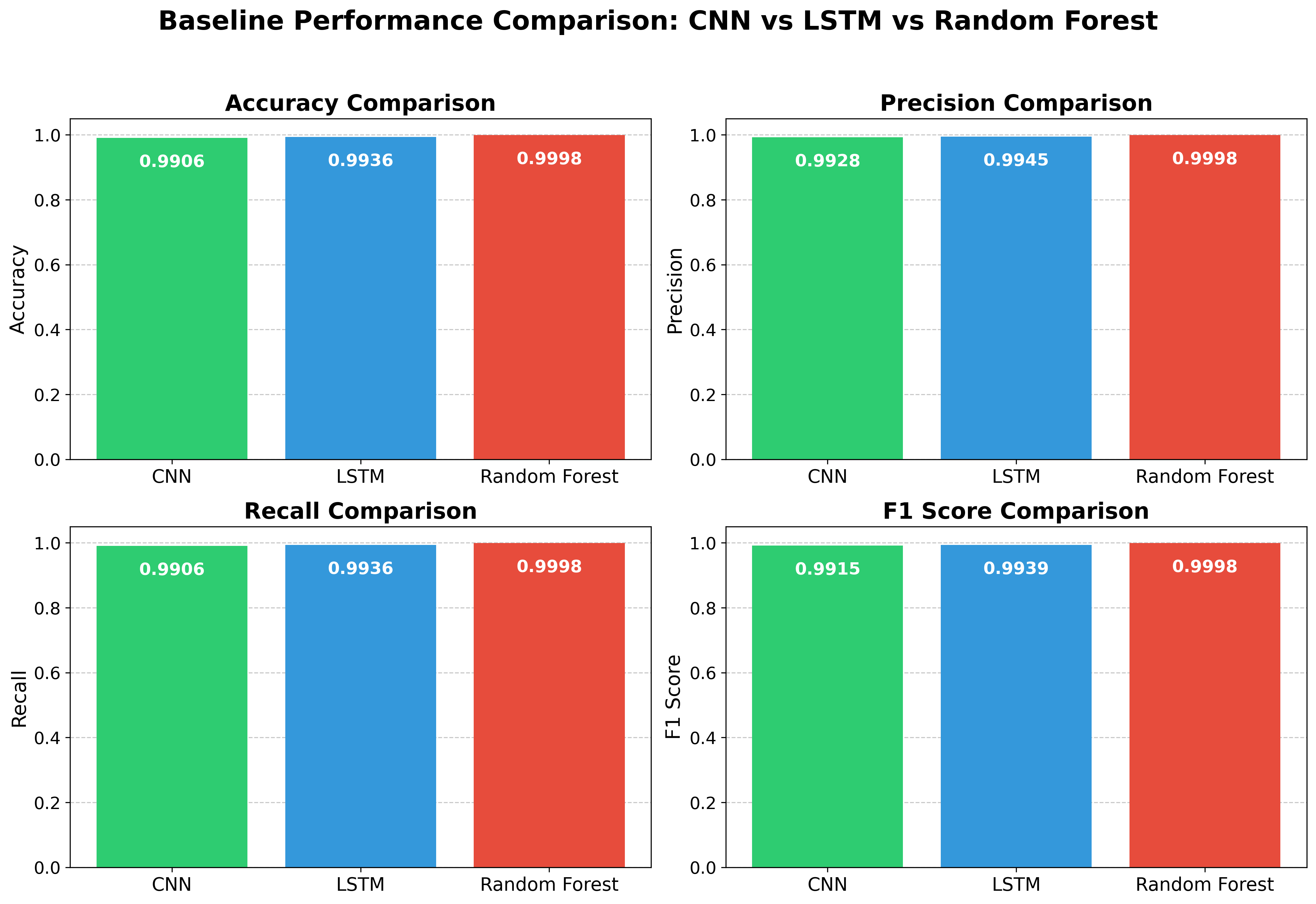}
\caption{Baseline performance comparison showing accuracy, precision, recall, and F1 scores for CNN, LSTM, and Random Forest on clean (non-adversarial) test data. All models achieve $>$99\% across all metrics, with Random Forest achieving near-perfect 99.98\%.}
\label{fig:baseline}
\end{figure}

Before throwing adversarial attacks at our models, we needed to confirm they could actually do their job well under normal conditions. Fig.~\ref{fig:baseline} shows how all three models performed on clean test data, and honestly, the numbers are impressive across the board. Random Forest came out on top with near-perfect scores: 99.98\% accuracy, 99.98\% precision, 99.98\% recall, and 99.98\% F1. LSTM was not far behind at 99.36\% accuracy and 99.39\% F1, while CNN achieved 99.06\% accuracy and 99.15\% F1. Table~\ref{tab:comprehensive_results} summarizes these baseline metrics alongside the adversarial results we will discuss shortly. These baseline results establish the reference point required to evaluate RQ2, which examines whether high clean accuracy reliably predicts adversarial robustness.

\begin{figure}[t]
\centering
\includegraphics[width=\columnwidth]{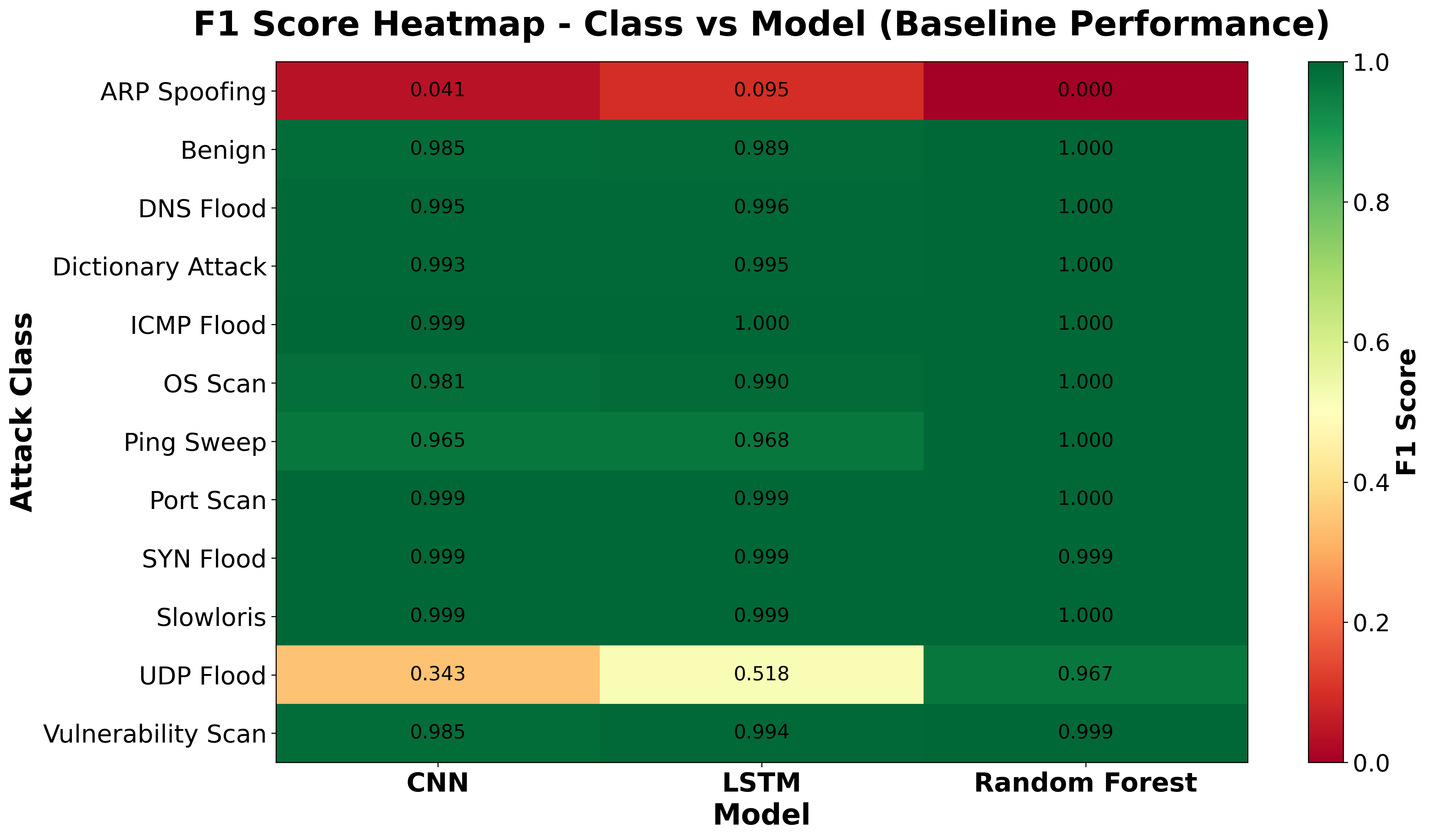}
\caption{Class-wise baseline F1 scores across all models. Green indicates strong performance ($>$0.9), yellow indicates moderate performance, and red indicates poor performance ($<$0.2). ARP Spoofing fails across all models due to extreme class imbalance (only 6 samples). UDP Flood shows weakness in neural networks but not Random Forest.}
\label{fig:classwise_heatmap}
\end{figure}

We would like to point out that these aggregate numbers do not tell the whole story. When we broke things down by attack class (Fig.~\ref{fig:classwise_heatmap}), some clear patterns emerged. All three models struggled with ARP Spoofing: CNN managed only 4.1\% F1, LSTM hit 9.5\%, and Random Forest completely failed at 0.0\%. UDP Flood detection was another problem, with CNN at 34.3\% and LSTM at 51.8\%, though Random Forest handled it reasonably well at 96.7\%.

These class-level weaknesses make sense when you consider the data distribution. ARP Spoofing had just 6 samples in the entire dataset, so there simply was not enough training data for any model to learn meaningful patterns. UDP Flood, while more common, apparently has feature characteristics that overlap with other traffic types, making it harder to distinguish.

While Random Forest achieves the highest baseline performance, subsequent adversarial evaluations reveal that baseline accuracy alone is insufficient for assessing robustness.

\subsection{Adversarial Attack Results}
\label{sub:E-B}

This subsection addresses RQ1 and RQ2 by evaluating how different model architectures respond to adversarial perturbations and whether baseline performance correlates with adversarial robustness.

We test each model with FGSM and PGD attacks at three perturbation levels ($\epsilon \in \{0.01, 0.05, 0.1\}$), and the results upended our initial expectations about which architecture would prove most robust.

\begin{figure}[t]
\centering
\includegraphics[
    width=1.00\columnwidth,
    height=0.55\textheight,
    keepaspectratio
]{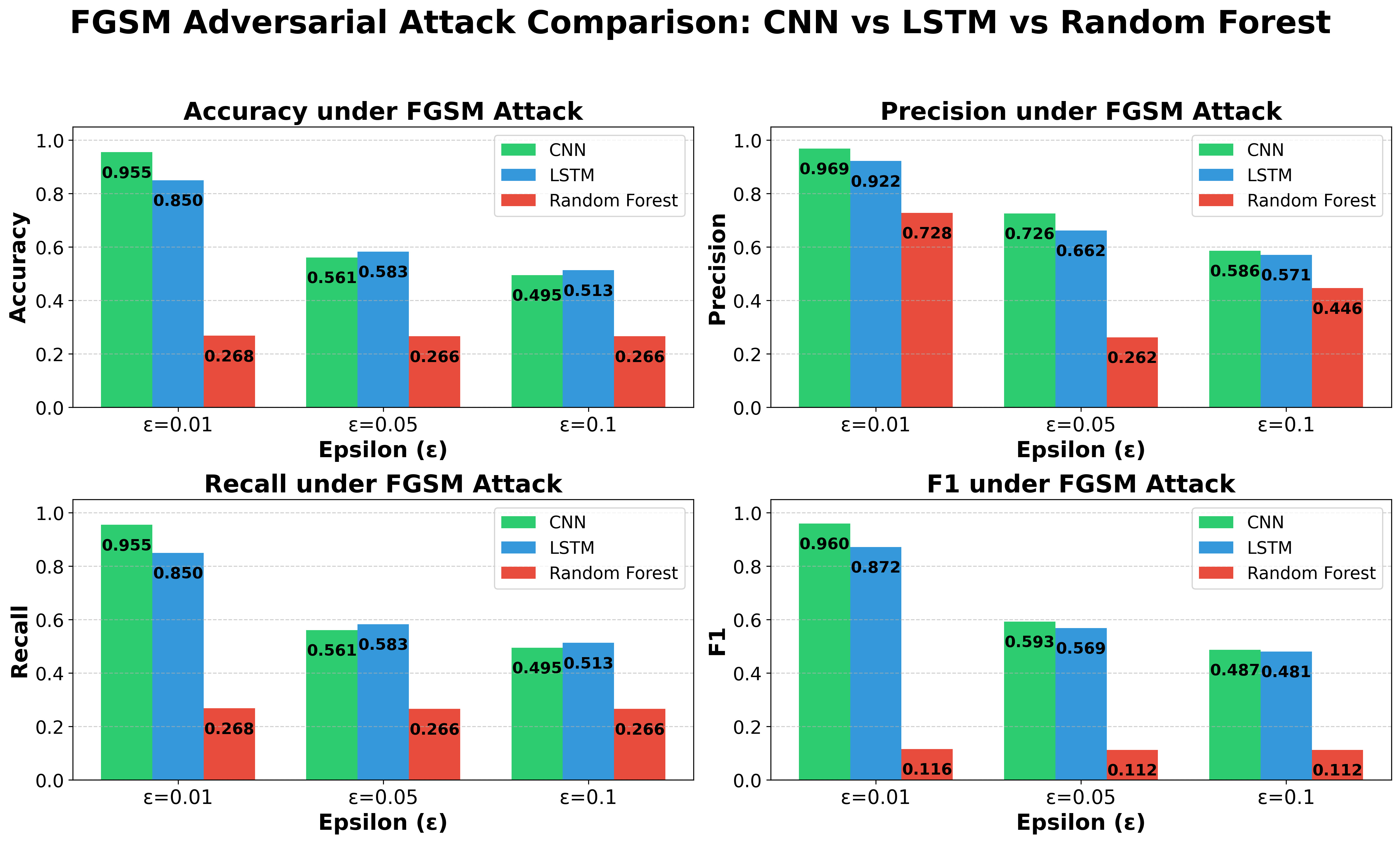}
\caption{Performance under FGSM adversarial attacks at $\epsilon \in \{0.01, 0.05, 0.1\}$. Random Forest (red) collapses immediately at $\epsilon=0.01$, dropping from 99.98\% to 26.8\% accuracy. CNN (green) and LSTM (blue) degrade gradually, with CNN showing superior robustness at low perturbation levels.}
\label{fig:fgsm}
\end{figure}

\begin{figure}[t]
\centering
\includegraphics[width=1.00\columnwidth,
    height=0.55\textheight,
    keepaspectratio]{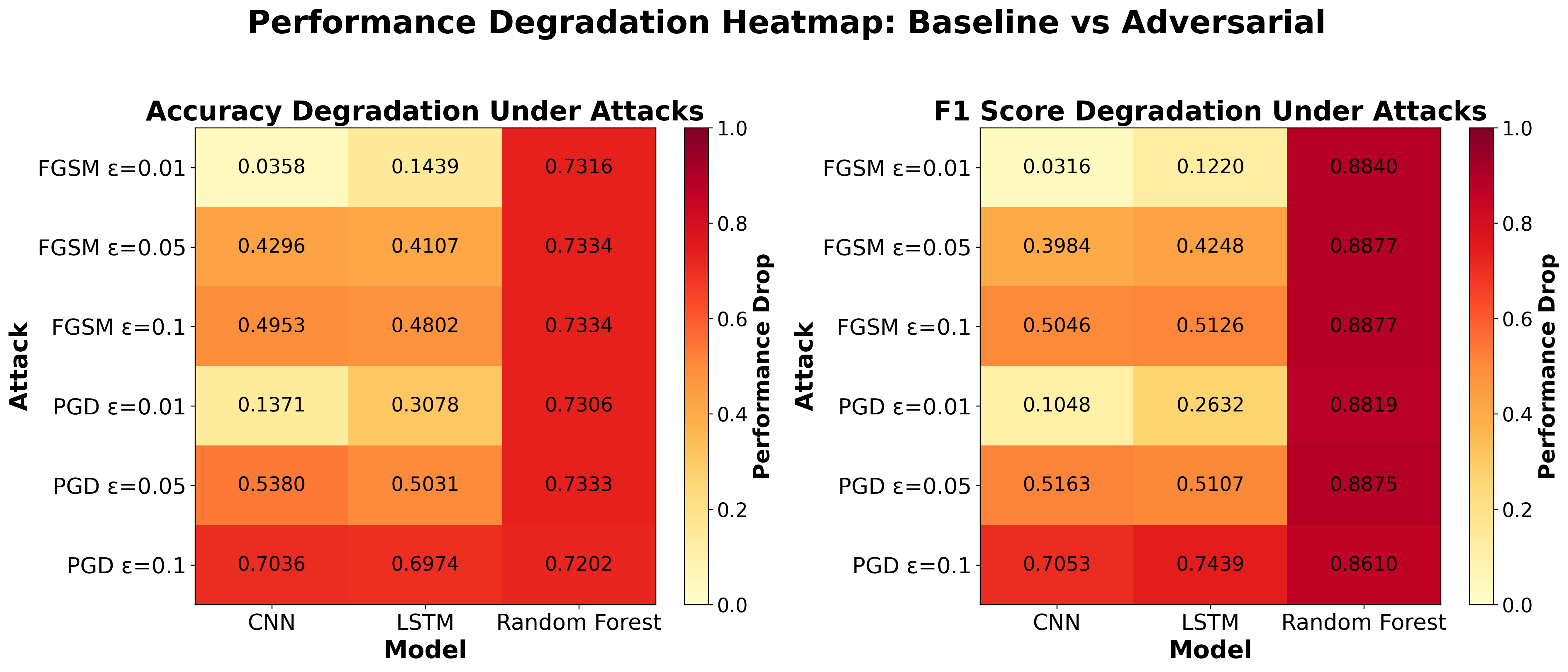}
\caption{Performance degradation heatmap comparing accuracy and F1 score drops from baseline under all attack configurations. Darker red indicates greater degradation. Random Forest shows consistently severe degradation ($>$70\% accuracy drop, $>$85\% F1 drop) across all attacks, while neural networks show proportional degradation scaling with $\epsilon$.}
\label{fig:degradation}
\end{figure}

Fig.~\ref{fig:fgsm} shows performance under FGSM attacks, and the first thing that jumps out is Random Forest's immediate collapse. Even at the smallest perturbation ($\epsilon = 0.01$), Random Forest's accuracy plummeted from 99.98\% to just 26.8\%, which is a 73.2 percentage point drop. Its F1 score fared even worse, crashing from 99.98\% to 11.6\%. Compare that to the neural networks. At $\epsilon = 0.01$, CNN retained 95.5\% accuracy (only 3.6\% degradation), and LSTM held onto 85.0\% (14.4\% degradation). Even at the most aggressive $\epsilon = 0.1$, CNN still achieved 49.5\% accuracy and 48.7\% F1, while LSTM managed 51.3\% accuracy and 48.1\% F1. The degradation heatmap (Fig.~\ref{fig:degradation}) makes this contrast stark. Random Forest shows uniformly high degradation (dark red) across all attack configurations, while CNN and LSTM show a more gradual progression from light to dark as $\epsilon$ increases.

\begin{figure}[t]
\centering
\includegraphics[
    width=\columnwidth,
    height=0.50\textheight,
    keepaspectratio
]{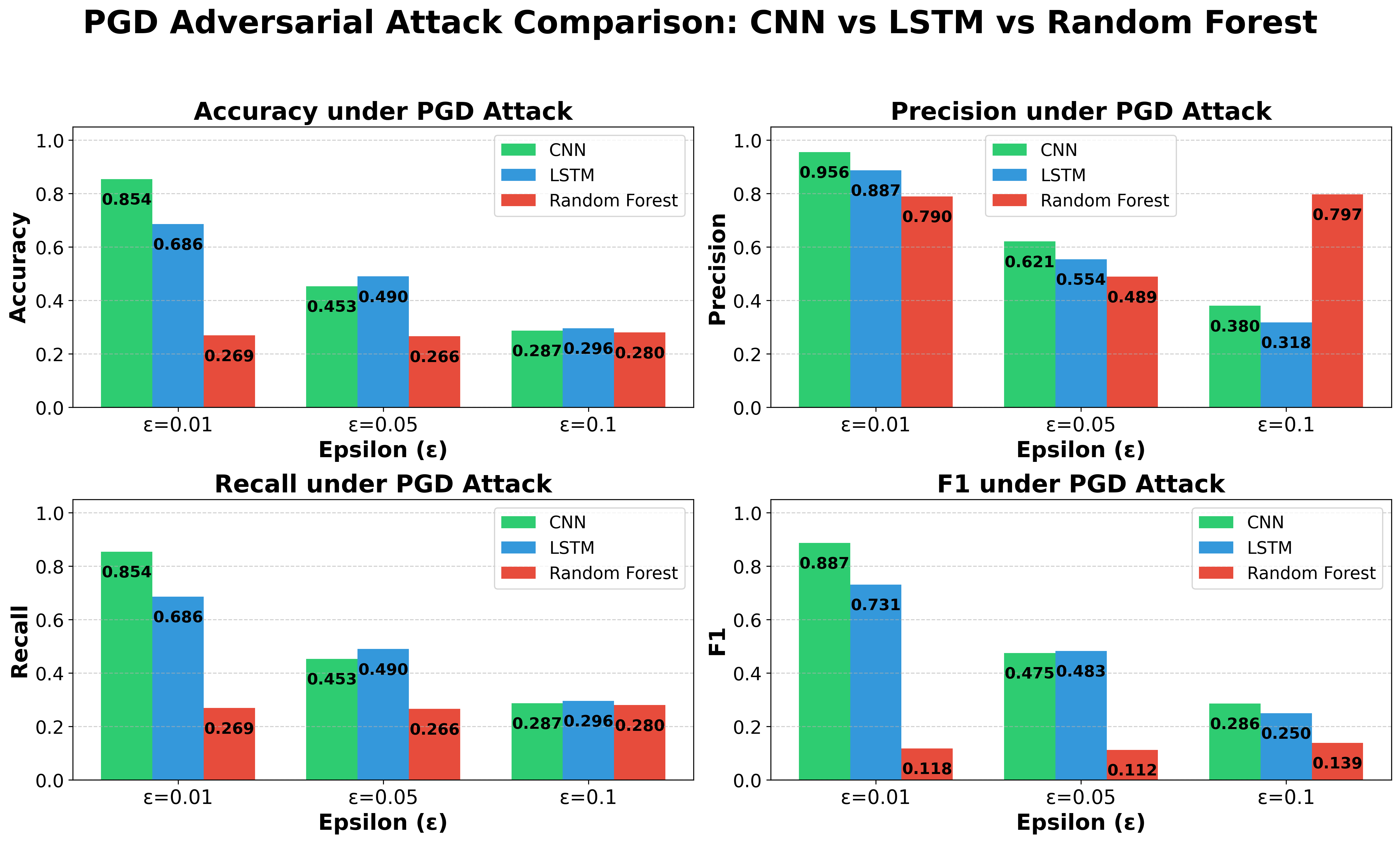}
\caption{Performance under PGD adversarial attacks at $\epsilon \in \{0.01, 0.05, 0.1\}$ with 40 iterations and step size $\alpha=\epsilon/4$.}
\label{fig:pgd}
\end{figure}

PGD, being an iterative attack, should theoretically be stronger than single-step FGSM. And for the neural networks, it was. However, the pattern of Random Forest's vulnerability persisted. Fig.~\ref{fig:pgd} breaks down the PGD results. At $\epsilon = 0.01$, PGD hit LSTM harder than FGSM did (68.6\% accuracy vs. 85.0\%), and CNN dropped to 85.4\% (vs. 95.5\% under FGSM). But Random Forest was already so degraded under FGSM that PGD could not make things much worse at 26.9\% accuracy, essentially unchanged from its FGSM performance. At $\epsilon = 0.1$, all models converged to roughly similar accuracy levels: CNN at 28.7\%, LSTM at 29.6\%, and Random Forest at 28.0\%. However, F1 scores told a different story: CNN achieved 28.6\%, LSTM got 25.0\%, but Random Forest collapsed to just 13.9\%. This suggests Random Forest was not just misclassifying samples; it was making systematically worse predictions.

\begin{table*}[!t]
\centering
\caption{Class-wise Baseline F1 Scores Across All Models with Attack Category Mapping}
\label{tab:classwise_f1}
\renewcommand{\arraystretch}{1.1}
\small
\begin{tabular}{l|ccc|l|r}
\toprule
\textbf{Attack Class} & \textbf{CNN} & \textbf{LSTM} & \textbf{Random Forest} & \textbf{Category} & \textbf{Test Samples} \\
\midrule
ARP Spoofing & \cellcolor{red!30}0.041 & \cellcolor{red!30}0.095 & \cellcolor{red!40}0.000 & Spoofing & 1 \\
Benign & 0.985 & 0.989 & \textbf{1.000} & Benign & 66,299 \\
DNS Flood & 0.995 & 0.996 & \textbf{1.000} & DoS & 10,616 \\
Dictionary Attack & 0.993 & 0.995 & \textbf{1.000} & Brute Force & 7,418 \\
ICMP Flood & 0.999 & \textbf{1.000} & \textbf{1.000} & DoS & 30,996 \\
OS Scan & 0.981 & 0.990 & \textbf{1.000} & Recon & 4,753 \\
Ping Sweep & 0.965 & 0.968 & \textbf{1.000} & Recon & 9,196 \\
Port Scan & 0.999 & 0.999 & \textbf{1.000} & Recon & 88,270 \\
SYN Flood & 0.999 & 0.999 & 0.999 & DoS & 13,583 \\
Slowloris & 0.999 & 0.999 & \textbf{1.000} & DoS & 10,393 \\
UDP Flood & \cellcolor{yellow!40}0.343 & \cellcolor{yellow!40}0.518 & 0.967 & DoS & 142 \\
Vulnerability Scan & 0.985 & 0.994 & 0.999 & Recon & 4,616 \\
\midrule
\textbf{Weighted Average} & 0.992 & 0.994 & \textbf{1.000} & --- & 246,283 \\
\bottomrule
\end{tabular}
\vspace{2mm}
\begin{flushleft}
\footnotesize
\textit{Note:} Red cells indicate severe underperformance (F1 $<$ 0.1). Yellow cells indicate moderate weakness (F1 $<$ 0.6). Class imbalance is evident: Port Scan dominates with 88,270 test samples while ARP Spoofing has only 1. This imbalance directly correlates with model performance, showing minority classes getting consistently lower F1 scores.
\end{flushleft}
\end{table*}

\subsubsection{Attack Comparison \& Trends}

\begin{figure}[t]
\centering
\includegraphics[width=\columnwidth,
    height=0.50\textheight,
    keepaspectratio]{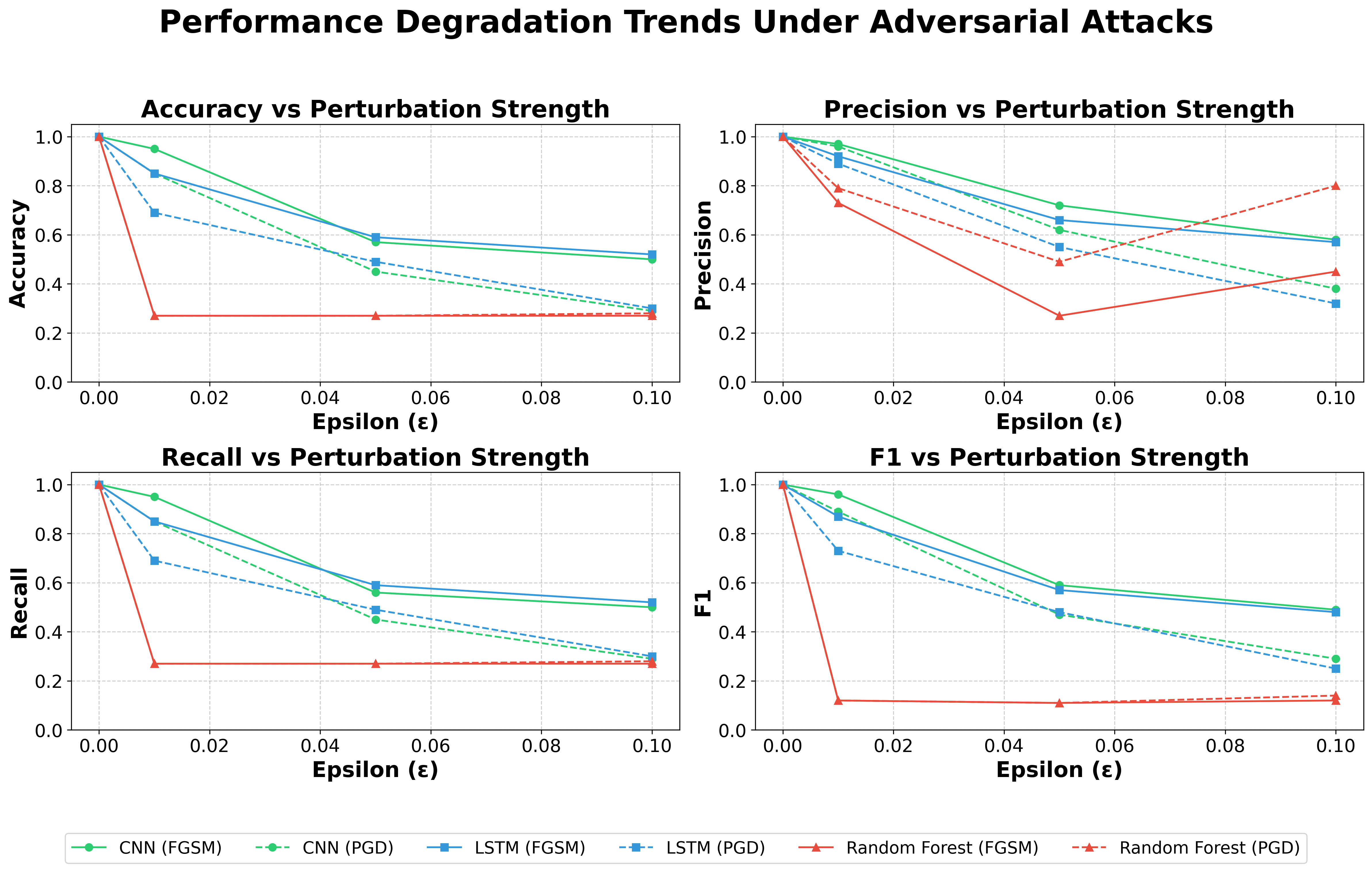}
\caption{Performance degradation trends across perturbation strengths. CNN and LSTM show gradual, proportional decline (smooth curves), while Random Forest drops vertically at $\epsilon=0.01$ then flatlines. Solid lines represent FGSM attacks; dashed lines represent PGD attacks. This pattern confirms Random Forest's brittle threshold-based decision boundaries.}
\label{fig:trends}
\end{figure}

Fig.~\ref{fig:trends} plots the degradation trajectories, and the contrast is striking. CNN and LSTM follow similar curves by starting high at baseline, declining gradually as $\epsilon$ increases, with PGD (dashed lines) consistently below FGSM (solid lines). Their degradation is roughly proportional to perturbation strength.

  Random Forest's trajectory is completely different. It drops almost vertically from baseline to its degraded state at $\epsilon = 0.01$, then it flatlines. Increasing $\epsilon$ from 0.01 to 0.1 barely changes its performance. This tells us that Random Forest's decision boundaries are very brittle that minimal perturbations are enough to cross them. Once the attack has done that, adding more perturbation is not needed because the damage is already done. These results provide direct evidence for RQ2, demonstrating that the model with the highest baseline accuracy can exhibit the most severe adversarial vulnerability.

\begin{figure*}[t]
\centering
\includegraphics[width=0.8\linewidth]{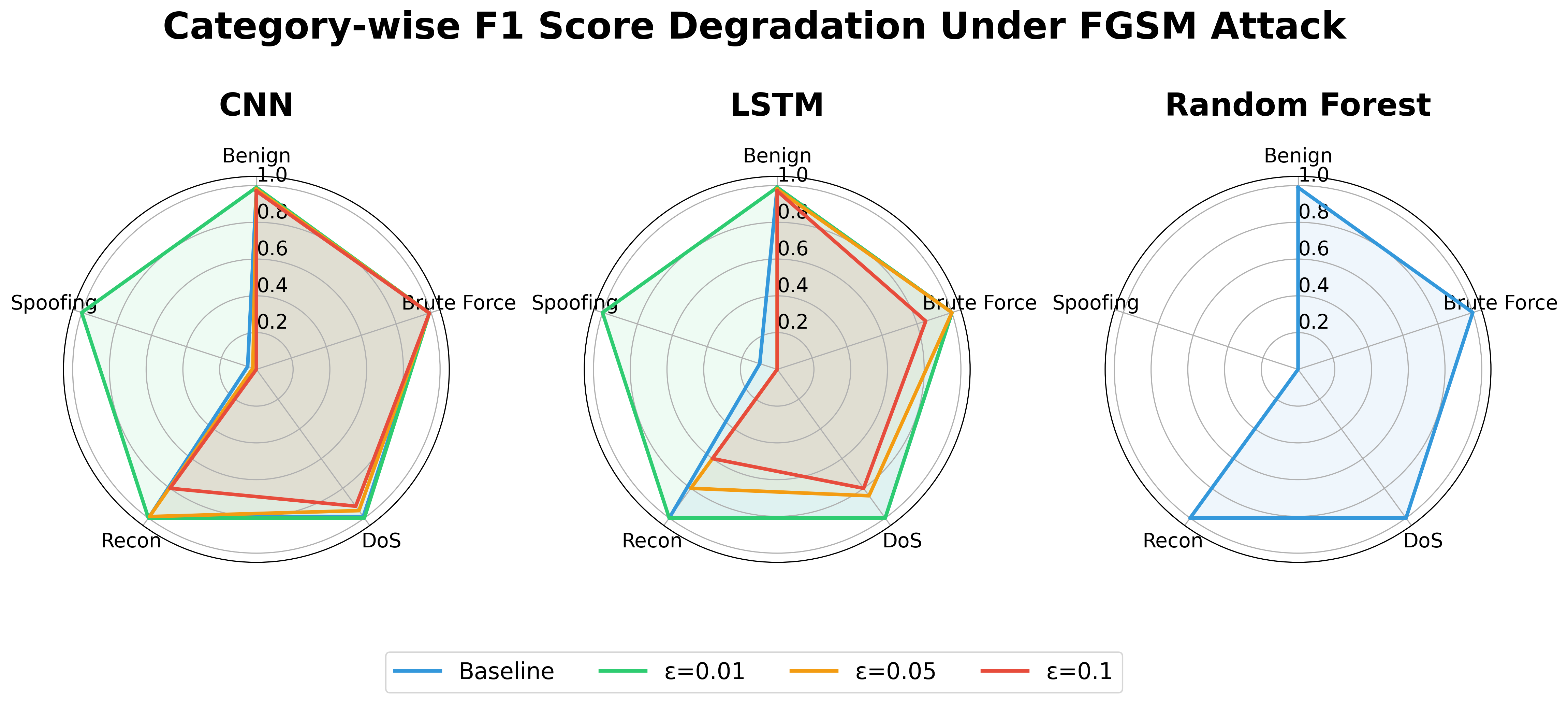}
\caption{Category-wise F1 score degradation under FGSM attacks for each model. CNN and LSTM show similar vulnerability patterns across categories, with Benign traffic most affected. Random Forest collapses uniformly across all categories at $\epsilon=0.01$, supporting our hypothesis that adversarial perturbations undermine its entire decision mechanism rather than exploiting specific category weaknesses.}
\label{fig:category_radar}
\end{figure*}

  Looking at the five attack categories (Benign, Reconnaissance, DoS, Brute Force, Spoofing) revealed which types of attacks became undetectable under adversarial perturbation. Fig.~\ref{fig:category_radar} shows the category-wise F1 degradation under FGSM. For CNN and LSTM, the vulnerability patterns were similar across categories. Benign traffic classification suffered the most, dropping from near-perfect to around 0.2 F1 at $\epsilon = 0.1$. Reconnaissance and DoS categories showed moderate degradation, while Brute Force remained relatively stable even under attack.

\begin{figure*}[t]
\centering
\includegraphics[width=0.8\linewidth]{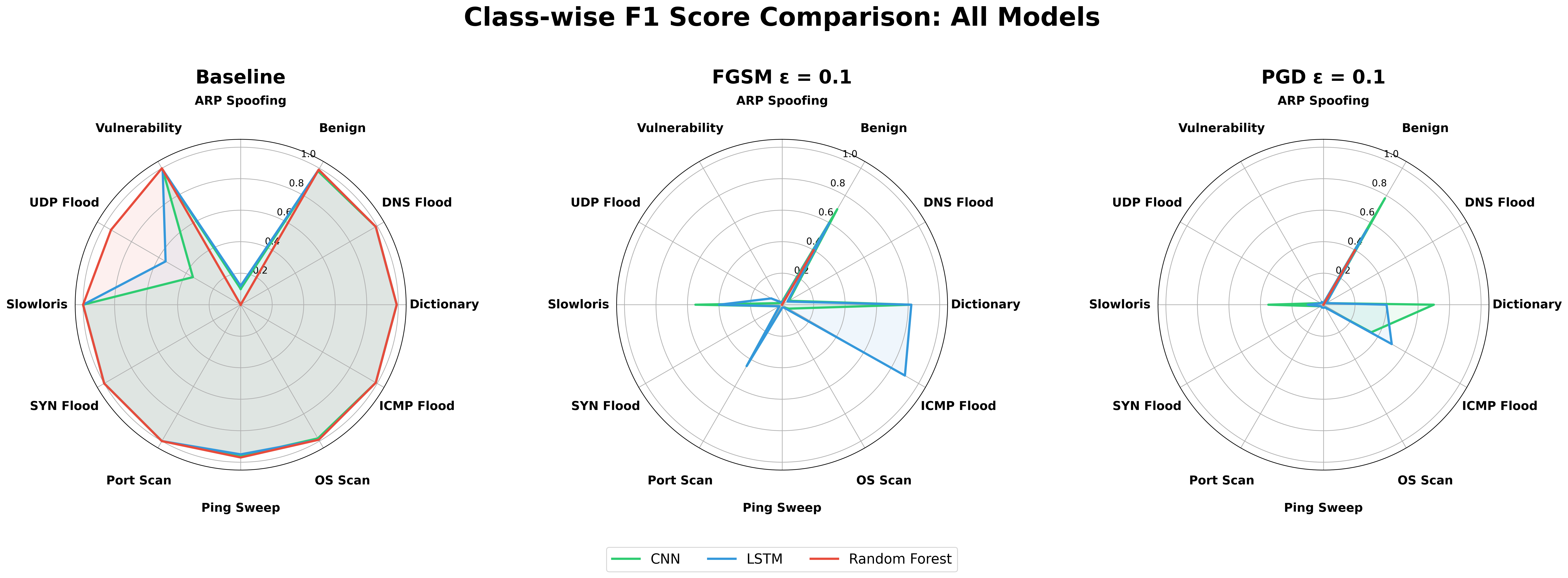}
\caption{Class-wise F1 score radar charts comparing baseline (left), FGSM $\epsilon=0.1$ (center), and PGD $\epsilon=0.1$ (right). At baseline, all models show full polygons except for ARP Spoofing and UDP Flood weak spots. Under adversarial attack, CNN and LSTM contract but retain structure, while Random Forest collapses toward the center across nearly all classes.}
\label{fig:radar}
\end{figure*}

Random Forest's category-wise performance collapsed uniformly. At $\epsilon = 0.01$, every category except Brute Force dropped below 0.4 F1. Benign classification essentially failed completely. This uniform collapse supports our hypothesis about threshold-based decision boundaries where adversarial perturbations do not exploit specific category weaknesses but rather undermine the entire decision-making mechanism. The radar charts in Fig.~\ref{fig:radar} provide a visual summary of class-wise robustness. At baseline, all three models show relatively full radar polygons (except for the known weak spots in ARP Spoofing and UDP Flood). Under FGSM $\epsilon = 0.1$, the polygons contract dramatically.

CNN's polygon contracts too, but maintains some structure in DNS Flood, ICMP Flood, and several reconnaissance attacks retain partially  detectable. LSTM shows a similar pattern. Random Forest's polygon, however, collapses almost entirely toward the center, with most classes dropping to near-zero F1. Under PGD $\epsilon = 0.1$, CNN and LSTM show even more contraction, but interesting asymmetries emerge. LSTM retains better performance on Benign traffic, while CNN holds up better on OS Scan and Vulnerability Scan. These differential vulnerabilities suggest the models learned different feature representations, making them potentially complementary for ensemble defense strategies.

\subsection{Key Findings Summary}
\label{sub:E-C}

The following findings summarize the answers to the three research questions posed in this study.

\subsubsection{RQ1: How does adversarial robustness differ across fundamentally distinct ML architectures used in network intrusion detection systems?} 

Despite achieving the highest baseline performance (99.98\% across all metrics), Random Forest proved catastrophically vulnerable to adversarial perturbations. 
\subsubsection{RQ2: Is high baseline detection accuracy a reliable indicator of adversarial robustness in ML-based NIDS?}

Even minimal $\epsilon = 0.01$ perturbations caused 73\% accuracy drops and 88\% F1 drops. This confirms our hypothesis that axis-aligned decision boundaries create exploitable discontinuities.
\subsubsection{RQ3: How does adversarial vulnerability vary across attack categories and individual intrusion classes in NIDS?}

Minority classes (ARP Spoofing, UDP Flood) were vulnerable even at baseline and became completely undetectable under adversarial attack. Class-balanced training alone is insufficient when the underlying data is extremely scarce. 

Both CNN and LSTM showed proportional degradation, where performance declined gradually as perturbation strength increased. At $\epsilon = 0.01$, CNN retained 96\% of its baseline accuracy. This smooth degradation profile is more suitable for security applications where graceful failure is preferable to catastrophic collapse. Additionally the adversarial examples we crafted against CNN transferred effectively to Random Forest (and would likely transfer to any model trained on the same features). This has serious implications for black-box attack scenarios.

While LSTM occasionally outperformed CNN at specific $\epsilon$ values, CNN showed more consistent robustness across attack types and perturbation levels. For security-critical deployments, CNN's predictable degradation profile is advantageous.

\section{CONCLUSIONS AND FUTURE WORK}
\label{sec:5conc}
In this section, we will translate our experimental findings into practical deployment guidance, acknowledge where our study falls short, and point toward research directions that seem worth pursuing.

\subsection{Deployment Recommendations}

After running all these experiments, we would like to give practitioners something concrete to work with. Table~\ref{tab:deployment} presents our findings into scenario-specific recommendations essentially a cheat sheet for deciding which model to deploy based on your threat environment.

\begin{table}[t]
\centering
\caption{Deployment Recommendations Based on Adversarial Robustness Evaluation}
\label{tab:deployment}
\renewcommand{\arraystretch}{1.3}
\setlength{\tabcolsep}{3pt}
\footnotesize
\begin{tabular}{@{}p{2.2cm}|c|c|c|p{2.2cm}@{}}
\hline
\rowcolor{headerblue}
\textcolor{white}{\textbf{Scenario}} & \textcolor{white}{\textbf{CNN}} & \textcolor{white}{\textbf{LSTM}} & \textcolor{white}{\textbf{RF}} & \textcolor{white}{\textbf{Rec.}} \\
\hline
Early Detection \newline ($\epsilon \leq 0.01$) & \cellcolor{lightgreen}$\checkmark$ & \cellcolor{lightgreen}$\checkmark$ & \cellcolor{lightred}$\times$ & \textbf{CNN} \\
\hline
Heavy Attack \newline ($\epsilon = 0.1$) & \cellcolor{lightgreen}$\checkmark$ & \cellcolor{lightgreen}$\checkmark$ & \cellcolor{lightred}$\times$ & \textbf{CNN/ LSTM} \\
\hline
Benign Class. & \cellcolor{lightgreen}$\checkmark$ & \cellcolor{lightgreen}$\checkmark$ & \cellcolor{lightred}$\times$ & \textbf{CNN} \\
\hline
Recon Detection & \cellcolor{lightyellow}$\triangle$ & \cellcolor{lightyellow}$\triangle$ & \cellcolor{lightred}$\times$ & \textit{Needs AT} \\
\hline
DoS Detection & \cellcolor{lightgreen}$\checkmark$ & \cellcolor{lightgreen}$\checkmark$ & \cellcolor{lightred}$\times$ & \textbf{CNN/ LSTM} \\
\hline
Distribution Shift & \cellcolor{lightgreen}$\checkmark$ & \cellcolor{lightgreen}$\checkmark$ & \cellcolor{lightred}$\times$ & \textbf{CNN}$^\star$ \\
\hline
Production NIDS & \cellcolor{lightgreen}$\checkmark$ & \cellcolor{lightgreen}$\checkmark$ & \cellcolor{lightred}$\times$ & \textbf{CNN} \\
\hline
\end{tabular}
\vspace{1mm}
\begin{flushleft}
\scriptsize
$\checkmark$=Suitable, $\triangle$=Partial, $\times$=Not recommended, AT=Adversarial Training, $^\star$=Best overall
\end{flushleft}
\end{table}

Our experimental findings can translate directly into deployment guidance, though the recommendations vary considerably depending on threat model and operational context. For early-stage attack detection where adversaries have not yet refined their evasion techniques ($\epsilon \leq 0.01$), CNN emerges as the clear frontrunner. It maintains 95.5\% accuracy at this perturbation level, compared to 85.0\% for LSTM and a precipitous drop to 26.8\% for Random Forest. Security operations teams prioritizing early warning capabilities should find this gap decisive. As perturbations intensify toward $\epsilon = 0.1$, however, the neural networks begin converging where both CNN and LSTM settle around 50\% accuracy under FGSM and roughly 29\% under PGD. At these heavier attack strengths, either architecture serves adequately, though CNN's smoother degradation trajectory offers marginally better predictability.

The picture shifts when we consider specific detection tasks. Benign traffic classification, which is arguably the most consequential for day-to-day operations, since false positives drain analyst resources and erode organizational trust, favors CNN models, which sustains higher precision as perturbations escalate. Reconnaissance detection presents a mode difficult challenge, in which neither neural architecture handles adversarial perturbations to OS Scans, Port Scans, or Ping Sweeps particularly well. Deploying undefended models for reconnaissance detection in adversarial environments seems inadvisable. Adversarial training targeting these specific patterns would be a prerequisite. DoS attack detection presents more optimistic scenerio since they have inherently high-volume signatures in DNS Floods, ICMP Floods, SYN Floods, Slowloris. They remain detectable even under perturbation, and both CNN and LSTM perform reliably in this category.

Looking beyond immediate threat detection, distribution shift tolerance matters for any system expected to operate over extended periods. Since networks evolve continuously due to new devices, drifting traffic patterns and application updates. Our covariate and label shift experiments suggest CNN handles these transitions most gracefully, making it the stronger choice for deployments requiring longevity without constant retraining. Weighing all factors together such as baseline performance, adversarial robustness, degradation characteristics, and operational complexity, we recommend CNN-based architectures for production NIDS deployments operating in adversarial threat environments.

\subsection{Limitations}

We will acknowledge that we only used ACI-IoT-2023 dataset, which is solid and represents modern IoT attack traffic. Certain environments such as enterprise and cloud may look different. We still claim generalizibility due to the fact that we assessed architecture of the network rather than detection capability on the dataset. We understand the questions on if our findings generalize beyond this particular data distribution.

Second, our attacks assume white-box access. We had full knowledge of model architectures and could compute gradients directly. Real attackers might not have that luxury or they might use completely different approaches like query-based attacks or even physical-layer manipulation. The transfer attack results suggest our adversarial examples would still cause problems in black-box scenarios, but we did not test this exhaustively.

Third, we used fixed perturbation budgets. An adaptive adversary who adjusts their perturbations based on classifier feedback could potentially do more damage than our static FGSM and PGD attacks. However, this is irrelevant since all FGSM and PGD attacks will still result in broken network packet samples.

Fourth, we did not test any defenses such as adversarial training, input preprocessing, and certified robustness techniques. We wanted to establish baseline vulnerability first, but the natural next question is whether standard defenses would help.

Finally, computational constraints limited our hyperparameter exploration. Maybe a wider CNN or different activation functions would show better robustness. We tested reasonable architectures but did not optimize specifically for adversarial resilience.

\subsection{Future Research Directions}

Various research avenues emerge from these findings, each tackling the constraints of the present study and broadening its ramifications. The subsequent phase entails adversarial training specifically designed for network intrusion detection. Considering CNN's exceptional robustness, it is reasonable to formulate training methods centered on convolutional architectures. The difficulty involves creating adversarial traffic that retains the semantics of the attack. Perturbations must avoid detection while preserving the functional attributes characteristic of each attack type.
Our CLEVER score study yielded robustness estimates. Nevertheless, these estimates do not constitute guarantees. Methods such as randomized smoothing or interval bound propagation may produce formal certificates, enabling defenders to assert certifiable security claims instead of relying exclusively on empirical evaluation.

Our exploration indicates that CNN and LSTM have similar failure patterns.  The superior performance of LSTM in benign traffic identification and CNN in reconnaissance classification indicates that ensemble methods merit further exploration. Utilizing stacking topologies, applying voting mechanisms, or training adversarial ensembles could leverage these complementarities to attain robustness that neither model offers independently.
Features that retain discriminative power during perturbation would enhance any subsequent classifier.

In addition to model-level considerations, the threat model warrants more elaboration. Genuine intrusions manifest as temporal sequences of behaviors aligned with frameworks such as MITRE ATT\&CK, rather than as discrete packet classifications. Assessing resilience against consecutive attack chains would more accurately represent operational realities. Likewise, our assumption on the white-box attack, however conventional, diminishes the potential of adversarial capabilities. Evaluating against adaptive adversaries that iteratively query the NIDS through decision-based and score-based attacks would yield a more accurate threat assessment. Ultimately, actual implementation limitations remain unresolved. Production systems must meet latency requirements in conjunction with accuracy objectives, and comprehending the robustness-latency trade-off for CNN architectures is essential for practical applicability.


\section*{ACKNOWLEDGMENT}
This work was supported by the U.S. Military Academy (USMA) under Cooperative Agreement No. W911NF-22-2-0160. The views and conclusions expressed in this paper are those of the authors and do not reflect the official policy or position of the U.S. Military Academy, U.S. Army, U.S. Department of Homeland Security, or U.S. Government.

\bibliographystyle{IEEEtran}
\vspace{-1mm}
\bibliography{main}

\begin{IEEEbiography}[{\includegraphics[width=1in,height=1.25in,clip,keepaspectratio]{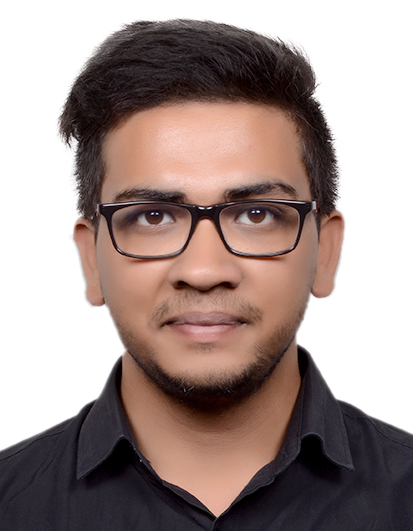}}]
{Mayank Raj}~(Student Member, IEEE \& COMSOC)~ is currently pursuing his Master of Science in Data Science with concentration in cybersecurity at the University of Massachusetts Dartmouth, where he serves as a Graduate Research Assistant under Dr. Gokhan Kul on a Department of Defense-funded cybersecurity project in collaboration with the U.S. Military Academy. He is a member of the IEEE Communications Society and serves as a reviewer for IEEE MILCOM 2025. His research interests include adversarial machine learning, network security, deep learning, and cybersecurity risk assessment.

\end{IEEEbiography}

\begin{IEEEbiography}
[{\includegraphics[width=1in,height=1.25in,clip,keepaspectratio]{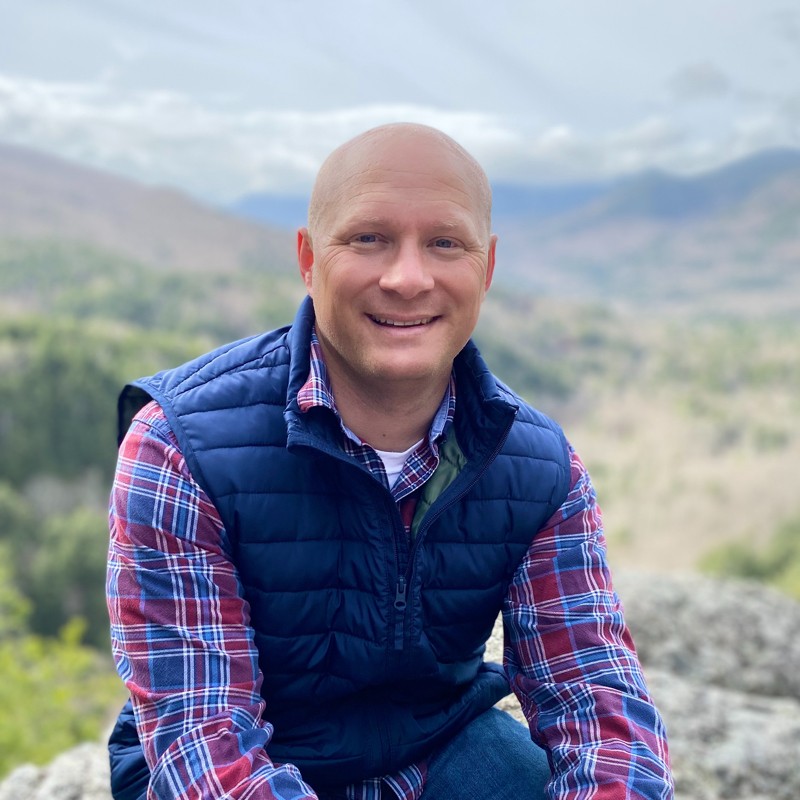}}]{Nathaniel D. Bastian}
(Senior Member, IEEE) received the Ph.D. degree in Industrial Engineering and Operations Research from Pennsylvania State University, University Park, PA, in 2016. He is currently an Assistant Professor in the Department of Electrical Engineering \& Computer Science at the United States Military Academy at West Point, and he serves as Deputy Director of the Robotics Research Center and Principal Investigator of the Laboratory for Artificial Intelligence Research \& Engineering (LAIRE). His primary research interests combine mathematical optimization, decision theory, machine learning, and statistical computing to design and develop secure, robust, and resilient AI-enabled autonomous C5ISRT systems. He has received \$8M in research funding support from DARPA, NSA, OUSD, DEVCOM, AFRL, ONR, etc.
\end{IEEEbiography}

\begin{IEEEbiography}[{\includegraphics[width=1in,height=1.25in,clip,keepaspectratio]{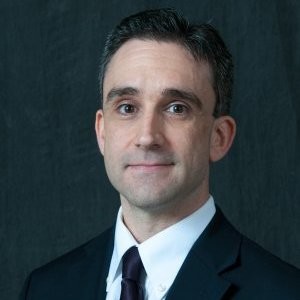}}]{Lance Fiondella}~(Member, IEEE)~received the Ph.D. degree in computer science \& engineering from the University of Connecticut, Storrs, CT, USA, in 2012. He is a professor of Electrical and Computer Engineering and Director of the Cybersecurity Center at the University of Massachusetts Dartmouth, an NSA/DHS designated Center of Academic Excellence in Cyber Research (CAE-R). His research has been funded by the Department of Homeland Security, NASA, the United States Department of Defense, and the National Science Foundation, including a CAREER Award and a CyberCorps Scholarship for Service (SFS) Award. He currently serves as an advisor to Working Group 35 (AI and Autonomous Systems) of the Military Operations Research Society. Previously, he served as associate editor of the Military Operations Research Journal and a technical committee chair of the Annual IEEE Symposium on Technologies for Homeland Security. He also served as the vice-chair of IEEE Standard 1633: Recommended Practice on Software Reliability from 2013-15 and a three-year term as a Member of the Administrative Committee of the IEEE Reliability Society from 2015-2017.
\end{IEEEbiography}

\begin{IEEEbiography}[{\includegraphics[width=1in,height=1.25in,clip,keepaspectratio]{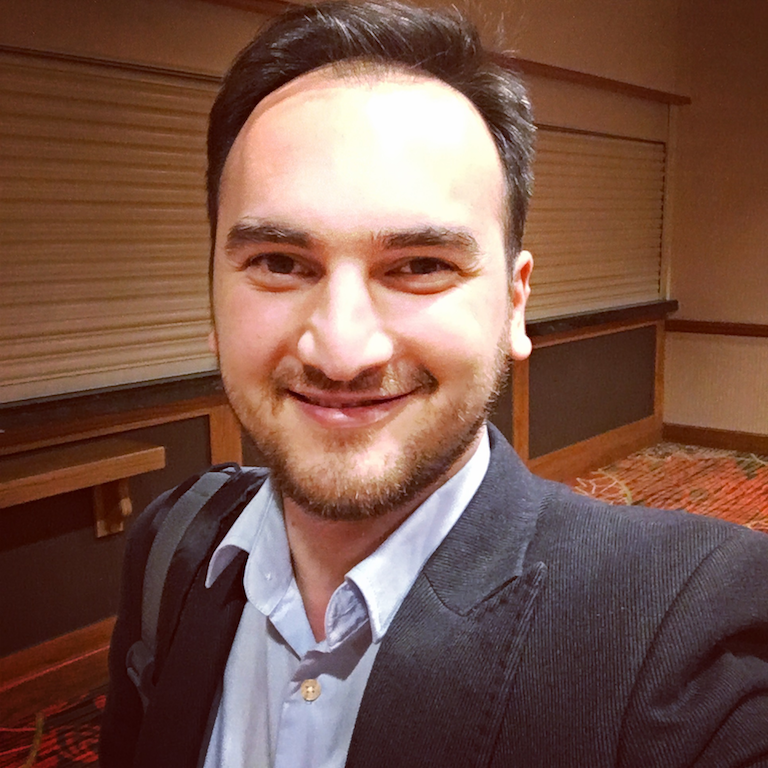}}]{Gökhan Kul}~(Member, IEEE)~received his B.S. degree in Computer Engineering from TOBB University of Economics and Technology, Ankara, Türkiye in 2010, his B.A. degree in Business Administration from Anadolu University, Eskisehir, Türkiye in 2012, his M.S. degree in Computer Engineering from Middle East Technical University, Ankara, Türkiye in 2012, and the Ph.D. degree in Computer Science from the University at Buffalo, SUNY in Buffalo, NY, USA, in 2018. 

He is an Associate Professor at the Department of Computer and Information Science and the Associate Director of the Cybersecurity Center at the University of Massachusetts Dartmouth. Prior to joining UMassD, he was an Assistant Professor at Delaware State University. He has authored or coauthored in reputable journals and conferences on subjects such as intrusion detection, data leakage, concept drift, threat detection, and software vulnerability assessment. His research focuses on AI for cybersecurity. Dr. Kul contributes to research reproducibility efforts as the Co-Chair of VLDB reproducibility program committee.
\end{IEEEbiography}

\EOD

\end{document}